\documentclass[11pt]{article}
\usepackage[margin=1in]{geometry}

\usepackage{xspace}
\usepackage{xcolor}
\usepackage{amscd}
\usepackage{amsmath}
\usepackage{amssymb}
\usepackage{amstext}
\usepackage{amsthm}
\usepackage{bm}
\usepackage{colonequals}
\usepackage{tikz}
\usepackage{booktabs}
\usepackage{array}
\usepackage{tabularx}
\usepackage{enumitem}
\usepackage[pagebackref=true,colorlinks,urlcolor=blue,linkcolor=purple,citecolor=blue,bookmarks,bookmarksopen,bookmarksnumbered]{hyperref}
\usepackage[capitalise,nameinlink]{cleveref}

\usepackage{mathtools}

\usepackage{array}
\usepackage{verbatim}
\usepackage{mathrsfs}
\usepackage{graphicx}
\usepackage{algorithm}
\usepackage{algpseudocode}
\usepackage{caption}



\newcommand*{\alglineref}[1]{\hyperref[#1]{Line~\ref*{#1}}}
\newcommand*{\stepref}[1]{\hyperref[#1]{Step~\ref*{#1}}}

\usepackage{tikz}
\usetikzlibrary{decorations.pathreplacing}

\usepackage[toc,page]{appendix}

\usepackage{titlesec}
\titleformat{\subsubsection}[runin]
{\normalfont\normalsize\bfseries}
{\thesubsubsection}{1em}{}

\theoremstyle{plain}
\newtheorem{theorem}{Theorem}[section]
\newtheorem{lemma}[theorem]{Lemma}

\newtheorem{corollary}[theorem]{Corollary}

\theoremstyle{definition}
\newtheorem{definition}[theorem]{Definition}

\newcommand{\N}{\mathbb{N}}

\DeclareSymbolFont{bbold}{U}{bbold}{m}{n}
\DeclareSymbolFontAlphabet{\mathbbold}{bbold}

\renewcommand{\epsilon}{\varepsilon}

\newcommand{\BPL}{\mathsf{BPL}}
\newcommand{\BPSPACE}{\mathsf{BPSPACE}}
\newcommand{\BPTISP}{\mathsf{BPTISP}}
\newcommand{\LocalRead}{\mathsf{LocalRead}}

\numberwithin{equation}{section}

\newif\ifshownotes
\shownotestrue

\ifshownotes
    \newcommand{\knote}[1]{{\color{cyan} (Kuan: #1)}}
	\newcommand{\hstnote}[1]{\textcolor{brown}{\textbf{[Shengtang: #1]}}}
    \newcommand{\wnote}[1]{\textcolor{purple}{\textbf{[Ruiyang: #1]}}}
\else
    \newcommand{\knote}[1]{}
	\newcommand{\hstnote}[1]{}
    \newcommand{\wnote}[1]{}
\fi

\title{Simulating Space Bounded Read-$k$ PTMs with Space Bounded Read-Once PTMs}
\author{
Kuan Cheng\thanks{CFCS, School of Computer Science, Peking University. ckkcdh@pku.edu.cn.}
\and
Shengtang Huang\thanks{CFCS, School of Computer Science, Peking University. peanuttang1320061044@gmail.com.}
\and
Ruiyang Wu\thanks{CFCS, School of Computer Science, Peking University. wuruiyang@stu.pku.edu.cn.}
}
\date{}

\begin{document}

\maketitle
\begin{abstract}
    We show that read-$k$ Probabilistic Turing Machines with bounded space $S$ and running time $T$ can be simulated by read-once Probabilistic Turing Machines with error $\varepsilon$, running in space $ O(k^2 (S + \log (n+ T)) + \log (1/\epsilon))$ and time $\frac{1}{\varepsilon} \cdot 2^{O\left(k^2\left(S+\log(n+T) \right)\right)}$. 
    As a direct consequence we show that read-$k$ $\mathsf{BPL}$ is equal to $\mathsf{BPL}$, for any constant $k$.
    Recently OpenAI announced that $\mathsf{BPL} = \mathsf{L}$. If this is true then our result immediately implies that read-$k$ $\mathsf{BPL}$ is equal to $\mathsf{L}$.
\end{abstract}

\section{Introduction}
Probabilistic Turing Machines (PTMs) with bounded space and bounded time are classic objects studied in computational complexity. The way of accessing randomness in such models may affect the computation power significantly. We mainly review the literature for bounded-error PTMs under two different settings of randomness accessing, with a special focus on the case of logspace machines.

The study of space-bounded randomness has developed along
two related directions: computation with fresh random bits,
equivalently read-once access to a random tape, and computation
with two-way access to a persistent random tape.
At logarithmic space and polynomial time, these models define
$\BPL$ and $\mathsf{BP}\cdot\mathsf{L}$, respectively.
Gill\cite{Gill1977} studied probabilistic
Turing machines whose transitions use coin tosses, corresponding
to the fresh-coin model.
Aleliunas, Karp, Lipton, Lov\'asz, and
Rackoff~\cite{AKLLR79} subsequently gave a randomized
logarithmic-space algorithm for undirected connectivity
using fresh random choices.
For arbitrary space-bounded machines in this read-once
randomness model, Borodin, Cook, and Pippenger~\cite{BCP1983}
obtained deterministic simulations using quadratic space.
The importance of the access convention was already apparent
in work of Karpinski and
Verbeek~\cite{KarpinskiVerbeek1985,KarpinskiVerbeek1986},
who studied two-way random tapes under space bounds without
a polynomial-time restriction and established limitations on
deterministic space simulation.
Their resource convention differs from the polynomial-time
two-way model defining $\mathsf{BP}\cdot\mathsf{L}$.
In the early 1990s, Nisan~\cite{Nisan1992PRG} constructed
a pseudorandom generator with $O(\log^2 n)$ seed length
that fools polynomial-time, logarithmic-space computations
with read-once access to randomness at inverse-polynomial
error.
Enumerating its seeds yields an $O(\log^2 n)$-space
deterministic simulation; Nisan's subsequent
work~\cite{Nisan1992SC} achieved polynomial time and
$O(\log^2 n)$ space simultaneously for the same read-once
source model.
In a separate comparison of randomness-access
models~\cite{Nisan1993}, he showed that every language in
$\BPL$ admits a zero-error logarithmic-space algorithm
with two-way access to randomness, and characterized
$\mathsf{BP}\cdot\mathsf{L}$ as almost-$\mathsf{L}$,
the languages decidable in logarithmic space relative to
almost every oracle.
For read-once source computations using relatively few
random bits, Nisan and Zuckerman~\cite{NisanZuckerman1996}
showed how to reduce $\operatorname{poly}(S)$ fresh random
bits to $O(S)$ random bits while retaining $O(S)$ space
and approximately preserving the output distribution.
Moving beyond read-once access, Impagliazzo, Nisan, and
Wigderson~\cite{INW1994} constructed pseudorandom generators
for machines whose two-way random-tape head visits each
cell only constantly many times.
Their result gives $O(\log^2 n)$ seed length for this
bounded-local-rereading model at logarithmic space,
polynomial time, and inverse-polynomial error.
For read-once randomized computation, Saks and
Zhou~\cite{SaksZhou1999} improved the deterministic
space bound to $O(\log^{3/2}n)$, without a polynomial-time
guarantee.
Saks~\cite{Saks1996} surveyed the different conventions
for space-bounded randomness.
For the unrestricted two-way randomness model, Klivans
and van Melkebeek~\cite{KlivansVanMelkebeek2002} obtained
conditional derandomization through space-efficient
hardness-versus-randomness constructions: an exponential
circuit lower bound, at all sufficiently large input
lengths, for a function computable in deterministic
linear space implies
$\mathsf{BP}\cdot\mathsf{L}=\mathsf{L}$.
Gutfreund and Viola~\cite{GutfreundViola2004} gave
constructions yielding randomness-efficient error
reduction for $\mathsf{BP}\cdot\mathsf{L}$, where the
random description may be reread.
For the specific problem of undirected connectivity,
Reingold~\cite{Reingold05} obtained a deterministic
logarithmic-space algorithm.
Applications of two-way randomness appeared in many algorithmic applications. For example the work
of Flaxman and Frieze~\cite{FlaxmanFrieze2007} considers
a space-bounded computation which must reconstruct the same
randomly perturbed graph on subsequent accesses.
More recently, Hoza~\cite{Hoza2021} improved the
deterministic space bound for read-once randomized
computation to
$O(\log^{3/2}n/\sqrt{\log\log n})$, again without a
polynomial-time guarantee.
Bisht, Gupta, Nimbhorkar, and
Volkovich~\cite{BishtEtAl2024} placed arithmetic-formula
polynomial identity testing in
$\mathsf{BP}\cdot\mathsf{L}$, illustrating the role of
persistent randomness in reusing a random evaluation
point.
For read-once randomness, the recent preprint of
OpenAI~\cite{OpenAI2026} states a deterministic
logarithmic-space simulation together with additive
inverse-polynomial acceptance-probability approximation.
For two-way randomness with a constant local visit bound,
Cheng and Wu~\cite{ChengWu2026SC} obtained a deterministic
simulation using polynomial time and $O(\log^2 n)$ space,
with an inverse-polynomial additive error for the
acceptance probability.
There is a natural question behind the development of studying the two random access models: 
as the permitted number
of visits to each random-tape cell increases, when can
the resulting computation still be simulated using
read-once randomness without increasing its asymptotic
space bound?

Towards addressing this question, we focus on the read-multiplicity
model of Impagliazzo, Nisan, and Wigderson~\cite{INW1994}.
We use the following formal definition to describe such models. If we bound the number of visits for the cells in the random tape we come to the class of read-$k$ PTMs.
\begin{definition}[Locally Read-$k$ Machine]
\label{def:local-random-tape}
Let $k:\N\to\N_{\ge 1}$. A \emph{locally read-$k$ machine}
is a probabilistic Turing machine with a two-way read-only
input tape, a read--write work tape, and a read-only random
tape satisfying the following conditions.
\begin{itemize}
    \item \textbf{Local access to randomness.}
    The random-tape cells are indexed by $1,2,\ldots$ and
    contain independent uniformly random bits. The random
    head starts at cell $1$. At each step, it either stays
    at its current cell or moves to an adjacent cell,
    without moving left of cell $1$. Once the random-tape
    contents are fixed, the computation is deterministic.

    \item \textbf{Bounded visits.}
    A \emph{visit} to a random-tape cell is a maximal
    consecutive period during which the random head
    remains at that cell. The initial placement at cell
    $1$ begins its first visit. For every input of length
    $n$ and every assignment to the random tape, each
    random-tape cell is visited at most $k(n)$ times.
\end{itemize}
\end{definition}

\begin{definition}[$\LocalRead_k$-$\BPTISP(T,S)$]
\label{def:local-read-k-bptisp}
Let $k,T,S:\N\to\N_{\ge 1}$. The class $\LocalRead_k$-$\BPTISP(T,S)$ consists of all languages $L$ for which there exists a locally read-$k$ machine $M$ that halts within $O(T(n))$ time and uses $O(S(n))$ work space on every input of length $n$ and every random tape. Moreover, for every input $x$,
\begin{align*}
    x\in L &\implies \Pr_r[M(x;r)\text{ accepts}]\ge\frac23,\\
    x\notin L &\implies \Pr_r[M(x;r)\text{ accepts}]\le\frac13.
\end{align*}
\end{definition}
We write
\[
\LocalRead_k\text{-}\BPSPACE(S(n))
:=
\LocalRead_k\text{-}\BPTISP(2^{O(S(n))},S(n)).
\]
In particular,
\[
\LocalRead_k\text{-}\BPL
:=
\LocalRead_k\text{-}\BPSPACE(O(\log n)).
\]
An immediate question is whether locally read-$k$ randomized machines can
be simulated with read-once access to randomness while approximately preserving
their acceptance probabilities and attaining a small asymptotic space bound,
particularly when $k$ is constant.

\subsection{Our Results}
We show the following simulation of read-$k$ space bounded PTMs by read-once space bounded PTMs.
\begin{theorem}[Fresh-Coin Simulation]
\label{thm:local-read-simulation}
Let $k,T,S:\N\to\N_{\ge 1}$ be logspace constructible functions. Let $M$ be a locally read-$k$ machine running in time $O(T(n))$ and space $O(S(n))$ on every input and every random tape. For every $0<\varepsilon\le 1/2$, there exists a probabilistic TM $M'$ with a read-once random tape such that, on every input $x$ of length $n$, the output distributions of $M(x)$ and $M'(x)$ have total variation distance at most $\varepsilon$. Moreover, $M'$ runs in space
\[
O\left(k(n)^2\bigl(S(n)+\log(n+T(n))\bigr)+\log(1/\varepsilon)\right)
\]
and time
\[
\frac{1}{\varepsilon} \cdot 2^{O\left(k(n)^2(S(n)+\log(n+T(n)))\right)}
\]
on every random tape.
\end{theorem}
For constant $k$, this yields the following equality of
complexity classes.
\begin{corollary}
\label{cor:local-read-complexity}
For every constant integer $k\ge 1$ and every logspace constructible function $S(n)\ge\log n$,
\[
\LocalRead_k\text{-}\BPSPACE(S(n))
=
\BPSPACE(S(n)).
\]
In particular,
\[
\LocalRead_k\text{-}\BPL=\BPL.
\]
\end{corollary}

\subsection{Other Related Read-$k$ Models}
Several related models restrict repeated access in different
ways.
A \emph{$k$-pass random-tape machine} scans the same random
string at most $k$ times, typically resetting its random
head between successive left-to-right scans.
David, Papakonstantinou, and
Sidiropoulos~\cite{DavidPapakonstantinouSidiropoulos2011}
studied the applicability of Nisan's generator to such multipass models; Mandal, Pavan, and
Vinodchandran~\cite{MandalPavanVinodchandran2015}
investigated their derandomization, and Dixon, Mandal,
Pavan, and Vinodchandran~\cite{DixonEtAl2016}
studied their advice complexity.
We note here that since the local read-$k$ model can move their random tape heads freely, it can definitely simulate the $k'$-pass model, by setting $k= 2k'$.
However, before this work, it is highly unclear how to simulate local read-$k$ model by $k'$-pass models, as even for constant $k$, a straightforward simulation of read-$k$ models may need $\Theta(n)$ passes.
Another extensively studied model is the
\emph{read-$k$ oblivious branching program}.
Here the variable queried at each layer is fixed
independently of the input, and each variable occurs
at most $k$ times in this reading sequence.
The sequence may jump arbitrarily between variables;
there is no adjacent-cell head.
Gurjar and Volk~\cite{GurjarVolk2020} constructed
pseudorandom generators for read-$k$ oblivious branching
programs when the reading sequence is known to the
generator, and Doron and Lang~\cite{DoronLang2026}
subsequently improved the seed-length bounds.
Knowledge of the reading order is a separate issue even for
read-once programs, as illustrated by the unknown-order
generator of Forbes and Kelley~\cite{FK18}.
More general \emph{read-$k$ branching programs} allow
the next queried variable to depend on the current state.

\section{Technical Overview}
\label{subsec:technical-overview}

Let \(M\) be a locally read-\(k\) randomized Turing machine running in time \(T\) and space \(S\) on inputs in \(\{0,1\}^n\). For every \(0<\varepsilon\le1/2\), we will construct a randomized Turing machine \(M'\) with one-way access to its random tape. On inputs in \(\{0,1\}^n\), the machine \(M'\) runs in space
\[
O\!\left(k^2(S+\log(n+T))+\log(1/\varepsilon)\right)
\]
and time \(2^{O(k^2(S+\log(n+T)))}/\varepsilon\). For every \(x\in\{0,1\}^n\), the acceptance probabilities satisfy
\[
\left|\Pr[M'(x)\text{ accepts}]-\Pr[M(x)\text{ accepts}]\right|
\le\varepsilon.
\]

\paragraph{Decomposing executions using checkpoint transitions}
Fix an input \(x\). We may assume \(k\le T+1\). Let \(J_0=\{1,\ldots,T+1\}\) contain all random-tape positions that \(M(x)\) may read. A \emph{visit} is a maximal consecutive stay at a single random-tape cell. Because \(M\) is locally read-\(k\), once the head leaves a cell after that cell's \(k\)-th visit, it can never return to that cell. Consequently, the rest of the computation is confined to the side of the departed cell on which the head remains.

This observation lets us shrink the interval of accessible cells. Start with \(I=J_0\). In the remaining execution, consider the first departure that terminates the \(k\)-th visit to some cell \(v\in I\), and mark that departure. If the computation continues, replace \(I\) by the connected component of \(I\setminus\{v\}\) that contains the head immediately after this departure, and repeat the same rule. The marked departures therefore split the execution into a nested sequence of intervals, each obtained from the previous one by removing a single cell and keeping the side containing the head.

The retained interval alone does not determine how the simulation should continue, since some bits inside the interval may already have been read. We therefore also record the execution's \emph{boundary history}: the chronological list of crossings of the interval's boundary, together with their times, positions, and configurations (control state, input- and work-tape head positions, and work-tape contents). Given an assignment to the bits in the interval, we can reconstruct each portion of the past execution inside the interval from the recorded crossings that enter the interval, and we can check the reconstructed execution against the recorded crossings that leave it.

A \emph{checkpoint} \(D=(I_D,\mathcal H_D)\) consists of two objects. The interval \(I_D\) is the retained interval immediately after the marked departure, that is, the connected component of \(I\setminus\{v\}\) containing the head after it leaves \(v\). The boundary history \(\mathcal H_D\) is the chronological list of all crossings of the two boundaries of \(I_D\), from the start of the computation through the marked departure, together with their times, positions, and configurations. We form the checkpoint at this departure, before the head begins
its next visit. At that moment every cell in \(I_D\) has been visited fewer than \(k\) times: before the departure, every cell in \(I{\setminus\{v\}}\) had been visited fewer than \(k\) times, and \(I_D\subseteq I\setminus\{v\}\). Since each boundary of \(I_D\) is crossed \(O(k)\) times, the history \(\mathcal H_D\) has list size at most \(O(k)\). A \emph{checkpoint transition} is the passage from one checkpoint to the next, or from a checkpoint to the point where the computation stops.

\paragraph{Executions inside an interval}
To carry out the consistency checks recursively, we simulate
the computation inside an interval while using recorded
crossing information to skip the intervening computation outside it.
An \emph{execution constraint} \(\mathcal C\) on an interval
\(J\) specifies an initial configuration, random-head position,
and time; prescribed bit values and crossing histories; and
boundary tables specifying exits from and returns to \(J\).

For a fixed assignment \(u\in\{0,1\}^J\), let
\(\textbf{Exec}_r(\mathcal C,u)\) denote the stopping report
of the following deterministic execution.
Starting from the initial data specified by \(\mathcal C\),
the execution follows \(M\) inside \(J\), using \(u\) for
random reads and checking the requirements of \(\mathcal C\).
Each exit from \(J\) is handled by the boundary tables.
When the boundary checks pass and a return is supplied,
the execution skips the external computation and resumes
at the prescribed configuration and time.
The visit counts inside \(J\) are not reset when the execution
returns to \(J\).
An attempted \((r+1)\)-st visit causes overflow before
the new visit is counted or its bit is read.

The execution stops at the first normal termination,
overflow, or failed check.
Its \emph{stopping report} \(\rho\) records the configuration,
head position, time, and reason for stopping, together with
the details of a failed check when applicable.
The requirements of \(\mathcal C\) are checked during execution;
the assignment \(u\) need not satisfy them in advance.

\paragraph{Distributions of transitions between checkpoints }
Let \(2\le r\le k\), and let \(\mathcal C\) be an execution
constraint on an interval \(L\). For each assignment
\(u\in\{0,1\}^L\), let \(D_0,D_1,\ldots,D_m\) be the
successive checkpoints encountered by the constrained execution
under \(\mathcal C\) on \(u\) with visit threshold \(r\),
where the start of the execution is the initial checkpoint
\(D_0=(L,())\).
Let \(\rho=\textbf{Exec}_r(\mathcal C,u)\) be the corresponding
stopping report. The \emph{next outcome} after \(D_j\) is
\(O_j=D_{j+1}\) if \(j<m\), and \(O_m=\rho\).

Let \(U_L\) be uniform over \(\{0,1\}^L\).
Drawing \(u\) according to \(U_L\) induces a distribution
on checkpoint paths and stopping reports.
For a checkpoint \(D\) with
\(\Pr_{U_L}[D_j=D]>0\), define
\[
K_j^{\mathcal C,r}(D,o)
:=
\Pr_{U_L}\!\left[O_j=o\mid D_j=D\right].
\]
We call \(K_j^{\mathcal C,r}(D,\cdot)\) the
\emph{transition distribution} at \(D\).

\paragraph{The Markov property of checkpoints.}
For a checkpoint \(D\) with \(\Pr_{U_L}[D_j=D]>0\),
let \(I=I_D\).
Write \(U_I\) for the bits of \(U_L\) at positions in \(I\).
The following property is sufficient for sampling checkpoint
transitions without retaining the earlier checkpoints.
Suppose that, conditioned on \(D_j=D\), the internal
assignment \(U_I\) is uniform over the assignments consistent
with \(D\); that conditioning on earlier checkpoints
restricts only the bits of \(U_L\) outside \(I\); and that
the next outcome under \(\mathcal C\) depends only on
\(D\) and \(U_I\).
Then, for every sequence of checkpoints
\(d_0,\ldots,d_j\) with \(d_j=D\) that has
positive probability,
\begin{equation}\label{eq:overview-conditional}
\Pr_{U_L}\!\left[
O_j=o \mid D_0=d_0,\ldots,D_j=d_j
\right]
=K_j^{\mathcal C,r}(D,o).
\end{equation}
Under this property, for any stopping report \(\rho\), the chain rule of
conditional probability now gives
\begin{equation}\label{eq:overview-exec-decomposition}
\Pr_{U_L}\!\left[
\textbf{Exec}_r(\mathcal C,U_L)=\rho
\right]
=\sum_{m=0}^{T+1}\ \sum_{D_0,\ldots,D_m}
\left(\prod_{i=0}^{m-1}
K_i^{\mathcal C,r}(D_i,D_{i+1})\right)
K_m^{\mathcal C,r}(D_m,\rho).
\end{equation}
Here the inner sum ranges over checkpoint sequences of positive probability.
Thus, to reproduce the stopping distribution, it suffices
to sample the transition distributions \(K_j^{\mathcal C,r}\).

\paragraph{A recursive sampler for checkpoint transitions.}
We now replace direct simulation of each checkpoint transition
by a recursive rejection-sampling procedure which uses only read-once randomness.
At a current checkpoint, we first guess the next checkpoint
or stopping report, together with the information needed
to verify it. We then check this guess using a randomized
execution with a smaller visit bound. If the check fails,
we repeat both the guess and the check with fresh independent
randomness. The verification is itself randomized: a fixed
guess may pass for some assignments to the random tape and
fail for others. We will implement this verification by
recursively sampling the lower-level execution's report.

Specifically, write \(\textbf{Sampler}_r(\mathcal C)\) for
the proposed sampling procedure on a constraint \(\mathcal C\)
with interval \(L\). The desired identity is that, for every
stopping report \(\rho\),
\begin{equation}\label{eq:overview-sampler-law}
 \Pr[\textbf{Sampler}_r(\mathcal C)=\rho]
 =\Pr_{U_L}[\textbf{Exec}_r(\mathcal C,U_L)=\rho].
\end{equation}
For \(r=1\), follow \(\textbf{Exec}_1\), drawing a fresh
uniform bit when a position is first read and retaining
that bit throughout the visit. A second visit cannot read
the discarded bit: the execution returns overflow before
that read, unless an earlier check has already returned
a report. This gives the required report distribution.
For \(r\ge2\), initialize the current checkpoint to \(D_0\),
and consider a reachable current checkpoint \(D\), with
\(I=I_D\).

We call the guessed data a \emph{candidate} \(Y\).
It specifies a proposed next outcome \(\pi(Y)\), which is
either a checkpoint or a stopping report; the crossing
histories needed to verify a proposed new checkpoint;
and the information needed to complete an \(r\)-th visit,
if one occurs before that outcome.
The completion information is a pair \((v,\beta)\), where
\(v\in I\) is the position of this visit and \(\beta\) is
the guessed bit there, or a marker \(\bot\) indicating
that the next outcome occurs without a new \(r\)-th visit.

The visit bound decreases because, at \(D\), every position
in \(I\) has been visited fewer than \(r\) times.
Up to the first attempted \(r\)-th visit, the execution
inside \(I\) therefore obeys visit bound \(r-1\), counting
visits both before and after \(D\).
A capped-\((r-1)\) execution can reproduce this portion
and stop immediately before that visit. Its stopping
configuration, position, and time, together with the
proposed bit and the history counters, suffice to complete
the visit and determine the next outcome.
It must also verify the history through \(D\), since
reaching \(D\) already restricts the bits in \(I\).

Given \(Y\), the procedure
\(\textbf{CreateConstraint}(\mathcal C,D,Y)\) constructs
a child constraint \(\mathcal C_Y\) on \(I\) by combining
the inherited requirements of \(\mathcal C\), the histories
\(\mathcal H_D\), and the requirements specified by \(Y\).
If the construction returns \(\bot\), reject the trial.
Otherwise, the child execution is specified as follows.
The child retains the initial data of \(\mathcal C\) if
its initial position lies in \(I\). Otherwise, the
construction checks the first entrance into \(I\) recorded
in \(\mathcal H_D\), and starts the child immediately
after that entrance.
For a fixed assignment \(u\in\{0,1\}^I\), the execution
\(\textbf{Exec}_{r-1}(\mathcal C_Y,u)\) follows \(M\)
inside \(I\), checks the requirements of \(\mathcal C_Y\),
and uses its boundary tables for excursions outside \(I\).
It uses the same assignment \(u\) throughout and does not
reset visit counts at \(D\). It returns its first stopping
report; an attempted \(r\)-th visit produces an overflow
report before that visit is counted or its bit is read.

We can now specify the constraints for which we prove
\eqref{eq:overview-sampler-law}.
Let \(\mathcal C_0\) be the constraint on \(J_0\) with the
initial configuration of \(M(x)\), random-head position \(1\),
initial time \(0\), no prescribed bit values or crossing
histories, and empty boundary tables.
Set \(\mathfrak C_k=\{\mathcal C_0\}\).
For \(r=k,k-1,\ldots,2\), define \(\mathfrak C_{r-1}\)
to consist of all non-\(\bot\) outputs
\(\textbf{CreateConstraint}(\mathcal C,D,Y)\), where
\(\mathcal C\in\mathfrak C_r\), \(D\) is reached by
\(\textbf{Exec}_r(\mathcal C,u)\) for some assignment \(u\),
and \(Y\) ranges over all candidates for \(\mathcal C,D\).
This includes child constraints constructed from incorrect
guesses. The correctness proof establishes the checkpoint
property above for every \(2\le r\le k\) and
\(\mathcal C\in\mathfrak C_r\).

Given a child report \(\rho\), define the deterministic check
\[
 \chi_Y(\rho):=
 \textbf{FinishAndCheck}(\mathcal C,D,Y,\mathcal C_Y,\rho).
\]
This check first verifies that the child execution has
reproduced the history through \(D\).
If \(Y\) proposes \((v,\beta)\), it requires \(\rho\) to
be an overflow report immediately before an \(r\)-th visit
to \(v\). It recovers the history counters from the reported
time (using \Cref{lem:boundary-history-invariant}), completes
that visit using \(\beta\), and applies the prescribed
departure and boundary rules to obtain the next outcome.
It then compares this outcome with \(\pi(Y)\).
If \(Y\) proposes \(\bot\), the check instead requires
\(\rho\) itself to be the stopping report \(\pi(Y)\).
For a proposed new checkpoint, it also verifies the
crossing histories supplied by \(Y\).

We prove \eqref{eq:overview-sampler-law} and almost-sure
termination simultaneously for all
\(\mathcal C\in\mathfrak C_r\), by induction on \(r\).
The case \(r=1\) was described above.
For \(r\ge2\), fix \(\mathcal C\in\mathfrak C_r\) and assume
both assertions for every constraint in \(\mathfrak C_{r-1}\).
Whenever \(\mathcal C_Y\ne\bot\), the definition of the
families gives \(\mathcal C_Y\in\mathfrak C_{r-1}\),
including when \(Y\) is an incorrect guess.
The induction hypothesis therefore allows us to
sample the child report using read-once randomness and gives
\begin{equation}\label{eq:overview-exec-to-sample}
 \Pr[\chi_Y(\textbf{Sampler}_{r-1}(\mathcal C_Y))=1]
 =\Pr_{U_I}[\chi_Y(\textbf{Exec}_{r-1}(\mathcal C_Y,U_I))=1].
\end{equation}
Thus the recursive call followed by \(\chi_Y\) implements
the randomized verification of \(Y\): although \(\chi_Y\)
is deterministic given its report, that report is random.

Let \(\mathcal Y_r\) be the fixed space of candidate
encodings, and put \(Q_r=|\mathcal Y_r|\).
Each valid encoding represents one candidate, and each
candidate has a unique encoding. At \(D\), perform
independent trials of the following form:
\[
\begin{aligned}
 Y &\sim \operatorname{Unif}(\mathcal Y_r),\\
 \mathcal C_Y &\gets \textbf{CreateConstraint}(\mathcal C,D,Y),\\
 \rho &\gets \textbf{Sampler}_{r-1}(\mathcal C_Y),\\
 \mathsf{ok} &\gets \chi_Y(\rho).
\end{aligned}
\]
An invalid encoding or a construction returning \(\bot\)
rejects the trial before the recursive call.
Otherwise, accept precisely when \(\mathsf{ok}=1\).
The proposal and the recursive call use independent fresh
coins, and a rejected trial is repeated at the same \(D\).
For an accepted candidate, set \(o=\pi(Y)\).
If \(o\) is a checkpoint, update \(D\gets o\) and repeat
at that checkpoint; otherwise, return the stopping report
\(o\).

To determine the distribution of an accepted outcome,
consider first a fixed assignment \(u\) to \(I\).
If \(u\) is consistent with \(D\), exactly one candidate
with \(\mathcal C_Y\ne\bot\) satisfies
\[
 \chi_Y(\textbf{Exec}_{r-1}(\mathcal C_Y,u))=1.
\]
It is the candidate formed from a capped-\(r\) execution
under \(\mathcal C\) that reaches \(D\) with internal
assignment \(u\). Its next outcome, completion data, and
crossing histories do not depend on the compatible
assignment outside \(I\).
If \(u\) is inconsistent with \(D\), no candidate passes.
These assertions are proved in
\Cref{lem:canonical-candidate,lem:outcome-test}.
Thus the randomized check accepts each candidate with
probability equal to the probability that \(U_I\) is
consistent with \(D\) and produces that candidate.

Let \(q_D>0\) be the probability that \(U_I\) is consistent
with \(D\). Summing over candidates that propose outcome
\(o\) gives
\begin{equation}\label{eq:overview-kernel-exec}
 q_D K_j^{\mathcal C,r}(D,o)
 =\sum_{Y:\pi(Y)=o}
   \Pr_{U_I}[\chi_Y(\textbf{Exec}_{r-1}(\mathcal C_Y,U_I))=1],
\end{equation}
where candidates with \(\mathcal C_Y=\bot\) contribute
zero. Each assignment consistent with \(D\) and producing
\(o\) is counted in exactly one summand.
Combining this identity with
\eqref{eq:overview-exec-to-sample} and the uniform proposal
probability \(1/Q_r\), one trial satisfies
\begin{equation}\label{eq:overview-trial-law}
\begin{aligned}
 \Pr[\mathsf{ok}=1\land\pi(Y)=o]
 &=\frac{1}{Q_r}
   \sum_{Y:\pi(Y)=o}
   \Pr[\chi_Y(\textbf{Sampler}_{r-1}(\mathcal C_Y))=1]\\
 &=\frac{q_D}{Q_r}K_j^{\mathcal C,r}(D,o).
\end{aligned}
\end{equation}
Summing over \(o\), a trial succeeds with probability
\(q_D/Q_r>0\). Independent trials therefore accept almost
surely, and dividing \eqref{eq:overview-trial-law} by
\(q_D/Q_r\) shows that the first accepted outcome has
distribution \(K_j^{\mathcal C,r}(D,\cdot)\).

The sampler starts at the reachable checkpoint \(D_0\).
At each subsequent stage, the accepted outcome is supported
on actual next outcomes, so every checkpoint reached by the
sampler is reachable by the capped execution.
Each recursive call terminates almost surely by induction,
and each continuation strictly shrinks the current interval.
Hence the sampler also terminates almost surely.
By \eqref{eq:overview-conditional}, the same transition
distribution remains valid when conditioned on the entire
preceding checkpoint sequence. Consequently, the sampler
produces a checkpoint path \(D_0,\ldots,D_m\) followed by
report \(\rho\) with probability
\[
 \left(\prod_{i=0}^{m-1}
 K_i^{\mathcal C,r}(D_i,D_{i+1})\right)
 K_m^{\mathcal C,r}(D_m,\rho).
\]
This equals the corresponding summand in
\eqref{eq:overview-exec-decomposition}. Summing over paths
proves \eqref{eq:overview-sampler-law} and completes the
induction.

We now apply the sampler to the initial constraint \(\mathcal C_0\). By~\eqref{eq:overview-sampler-law} with \(r=k\) and \(\mathcal C=\mathcal C_0\), the procedure \(\textbf{Sampler}_k(\mathcal C_0)\) {has the same stopping-report distribution as \(M(x)\)}; in particular, its acceptance probability equals that of \(M(x)\). Put \(\mathfrak{b}=C_0(S+\log(n+T+2))\) for a sufficiently large constant \(C_0\). We define \(M'\) to run \(\textbf{Sampler}_k(\mathcal C_0)\) and, after \(2^{O(k^2\mathfrak{b})}/\varepsilon\) steps, to halt and return \(\bot\).

\paragraph{Space and running time.}
At recursion level \(r\), a boundary is crossed at most \(2r\) times, so a checkpoint or candidate uses \(O(r\mathfrak{b})\) bits, there are at most \(2^{O(r\mathfrak{b})}\) checkpoints per stage, and proposals can be generated with \(Q_r=2^{O(r\mathfrak{b})}\). Inherited history lists are stored by reference; their references, cutoffs, and counters, together with inherited bit requirements, use \(O(k\mathfrak{b})\) bits per level. Thus each level retains \(O((r+k)\mathfrak{b})\) bits, and the workspace is
\[
\sum_{r=1}^k O((r+k)\mathfrak{b})=O(k^2\mathfrak{b}).
\]

For the running time, a small \(q_D\) may require many trials, but it also makes \(D\) unlikely. If \(p_D\) is the probability of reaching \(D\) at its stage, consistency of the internal assignment is necessary, so \(p_D\le q_D\) and
\begin{equation}\label{eq:overview-time-cancellation}
p_D\frac{Q_r}{q_D}\le Q_r.
\end{equation}
Hence each checkpoint contributes at most \(Q_r\) times the cost of a trial to the expected cost, and with \(2^{O(r\mathfrak{b})}\) checkpoints per stage and \(T+1\) stages, level \(r\) costs \(2^{O(r\mathfrak{b})}\) times a trial in expectation. Unrolling over \(k\) levels yields expected time \(2^{O(k^2\mathfrak{b})}\). By Markov's inequality, truncating after \(2^{O(k^2\mathfrak{b})}/\varepsilon\) steps changes the acceptance probability by at most \(\varepsilon\). The cutoff counter uses \(O(k^2\mathfrak{b}+\log(1/\varepsilon))\) additional space, so \(M'\) runs in space \(O(k^2\mathfrak{b}+\log(1/\varepsilon))\) and time \(2^{O(k^2\mathfrak{b})}/\varepsilon\).

\section{Fresh-Coin Simulation}\label{sec:fresh-coin}

\paragraph{Some basic definitions}
A \emph{configuration} of $M$ consists of its input-tape head
position, finite control state, work-tape head position, and
work-tape contents. The random-tape head position is recorded
separately.
The initial configuration has time $0$. After each application
of $M$'s transition function, the time increases by $1$.
Thus, the configuration at time $t$ is the one reached after
exactly $t$ transitions.
For $q\ge1$, the \emph{cut} $q\mid q+1$ is the boundary between
random-tape positions $q$ and $q+1$. We also call it
\emph{cut $q$} for short. The random-tape head crosses this
cut when it moves from $q$ to $q+1$ or from $q+1$ to $q$.
We also use cut $0$ for the left boundary of cell $1$;
the random head never crosses this cut, so its history is empty.
For an interval $[a,b]$ of the random tape, we define
$\partial[a,b]=\{a-1,b\}, \partial_L [a,b] = a-1$ and $\partial_R [a,b]=b$.

A probability distribution on a finite set $\Omega$ is a
function $\mu:\Omega\to[0,1]$ satisfying
$\sum_{\omega\in\Omega}\mu(\omega)=1$.
We write $X\sim\mu$ when $X$ has distribution $\mu$, and
$\mathcal L(X)$ for the distribution of a random variable $X$.
The uniform distribution on $\Omega$ is denoted by
$\text{Unif}(\Omega)$. For a finite set $J$, the uniform
distribution on $\{0,1\}^J$ is denoted by $U_J$.

Fix an input $x\in\{0,1\}^n$ and a locally read-$k$ machine
$M$ using time at most $T$ and work space at most $S$.
All configurations below use at most $S$ work-tape cells.
When simulating a transition from any such configuration,
an attempted access beyond these cells instead enters a fixed
rejecting state, leaves all tape heads and work-tape contents
unchanged, and advances time by one. We use this convention
throughout, including in executions under constraints.
It does not change any computation of $M$ from its original
initial configuration, and ensures that simulations from
guessed configurations obey the same space bound.
All intervals below are contained in $\{1,\ldots,T+1\}$;
random-head positions in records and reports belong to
$\{1,\ldots,T+2\}$, allowing one step beyond the right endpoint.
All timestamps belong to $\{0,\ldots,T\}$.
Let
\[
 \mathfrak{b}:=C_0\bigl(S+\log(n+T+2)\bigr),
\]
where $C_0$ is a sufficiently large constant depending only on $M$. This budget suffices to encode a configuration, a random-head position, and a timestamp. 

We will consider recursive algorithms in this work. Each call to a recursive algorithm considered in this work
has a positive integer parameter $r$, called its \emph{rank}.
A call of rank $r\ge2$ may make recursive calls of rank $r-1$.
When a rank-$r$ call invokes a rank-$(r-1)$ call,
we refer to the former as the \emph{parent} and the latter
as the \emph{child}.
We also use these terms for their respective parameters and relative notions.

\paragraph{Organization of this section.}
In \Cref{subsec:sources}, we define capped-$r$ executions
under constraints and show that the initial constraint
reproduces the computation of $M$.
In \Cref{subsec:recursive-sampling}, we construct a recursive
sampler using read-once randomness.
We prove equality of its return-report distribution with
that of the corresponding capped execution in
\Cref{subsec:sample-correctness}.
Finally, \Cref{subsec:conclusion} bounds its space and expected
running time, and imposes a time cutoff to prove the main theorem.

\subsection{Capped-$r$ Executions under Constraints}
\label{subsec:sources}

We first define crossing histories, execution constraints,
return reports, and checkpoints, and then use them to define
capped-$r$ executions. Crossing histories record movements
across specified cuts; execution constraints prescribe the
initial data, bit and crossing requirements, and boundary
histories used by the execution.
\begin{definition}[Crossing History]
\label{def:crossing-history}
A \emph{crossing history} is a finite ordered sequence of records
\[
L=((s_j,x_j,y_j,c_j))_{j=1}^{\ell}.
\]
Here
$0\le s_1<\cdots<s_\ell<T$ and $x_j, y_j$ are indices of the random tape s.t. $|x_j-y_j|=1$. $c_j$ is a configuration of $M$ running on $x$.
One can think that a record $(s_j,x_j,y_j,c_j)$ means that the random-tape head is at position $x_j$ at time $s_j$, and that one transition moves it to position $y_j$ and produces configuration $c_j$ at time $s_j+1$. The empty history ($L$ being empty) is allowed.

A record $(s_j,x_j,y_j,c_j)$ crosses a cut $q \mid q+1$ if $\{x_j,y_j\}=\{q,q+1\}$.
If every record of $L$ crosses the cut $q \mid q+1$ , then say $L$ is a crossing history at the cut $q\mid q+1$.
\end{definition}

\begin{definition}[Execution Constraint]
\label{def:execution-constraint}
An \emph{execution constraint} is a tuple
\[
\mathcal C=
(J,\sigma_0,i_0,t_0,\mathcal H,\mathcal B,H_L,H_R),
\]
where:
\begin{itemize}
    \item $J=[a,b]\subseteq\{1,\ldots,T+1\}$ is a nonempty
    integer interval of random-tape positions.
    \item $(\sigma_0,i_0,t_0)$ specifies an initial configuration, random-head position, and a timestamp, with $i_0\in J$ and $0\le t_0\le T$.
    \item $\mathcal H=\{(q,L_q,\tau_q):q\in Q\}$ is a
    collection of \emph{crossing requirements}, where
    $Q\subseteq\{0,\ldots,T+1\}$ is a set of cut indices.
    Each entry has $t_0\le\tau_q\le T$, and $L_q$ is a
    crossing history at cut $q$ whose records all have
    timestamps in $[t_0,\tau_q)$.
    It prescribes the complete crossing history at that
    cut during this interval. Each cut has at most one
    entry in $\mathcal H$; cuts need not lie inside $J$.

    \item $\mathcal B$ is a finite collection of pairs $(v,\beta)\in J\times\{0,1\}$.
    
    One can think that $\mathcal B$ is used to remember that the bit at position $v$ of the random tape is $\beta$.
    \item $H_L,H_R$ are two crossing histories at $a-1\mid a$ and $b\mid b+1$, respectively, with the restriction that for any $(s_j, x_j, y_j, c_j)$ in $H_L$ or in $H_R$, the $s_j$ has to be in $[t_0,T)$. 
 
    Note that one can use $H_L,H_R$ to supply some information about those computations of $M$ that depend on the randomness out of $J$.
\end{itemize}
\end{definition}

A return report records where and why the execution stops,
together with the details of a failed check when applicable.
\begin{definition}[Return Report]
\label{def:return-report}
Given an execution constraint
\(
\mathcal C=(J,\sigma_0,i_0,t_0,\mathcal H,\mathcal B,H_L,H_R),
\)
a \emph{return report} is a tuple
\[
R=(\kappa,\sigma,i,t,\gamma),
\qquad
\kappa\in
\{\mathsf{HALT},\mathsf{OVERFLOW},\mathsf{FAIL}\}.
\]
Here $\sigma$ is a configuration, $i$ is a random-head
position, and $t$ is a timestamp.
The tag $\kappa$ will be used to indicate a normal termination, an overflow situation,
or a violation of a requirement of $\mathcal C$, respectively.

For $\kappa\in\{\mathsf{HALT},\mathsf{OVERFLOW}\}$,
we set $\gamma=\bot$.
For $\kappa=\mathsf{FAIL}$, using $\mathsf{history}, \mathsf{bit}, \mathsf{boundary}$ as 3 fixed labels, the field $\gamma$
has one of the following forms:
\begin{itemize}
    \item $(\mathsf{history},q,e)$, 
    identifying an entry
    $(q,L_q,\tau_q)\in\mathcal H$.
    Here $e$ is a crossing record involved in a
    disagreement with $L_q$, or a prescribed record
    of $L_q$ left unconsumed.

    \item $(\mathsf{bit},v,\beta)$, identifying a prescribed
    bit pair $(v,\beta)\in\mathcal B$ whose requirement
    is violated.

    \item $(\mathsf{boundary},s,e)$,  identifying a boundary
history $H_s$, where $s\in\{L,R\}$.
Here $e$ is an exit or entrance record involved in
the violation, with $e=\bot$ if a required entrance
record is absent.
\end{itemize}

For simplicity of description, we also use the terminology ``return $\kappa$ at $(\sigma, i, t)$ with $\gamma$'' to alternatively denote returning a report $(\kappa,\sigma,i,t,\gamma)$.

\end{definition}

A nonterminal checkpoint stores the current interval and its
boundary histories.
\begin{definition}[Nonterminal Checkpoint]
\label{def:checkpoint}
Fix an execution constraint $\mathcal C$ with interval $J$
and initial time $t_0$. A \emph{noninitial nonterminal checkpoint} is
\[
 D=(I_D,\mathcal H_D),
\]
such that
\begin{itemize}
    \item $I_D$ is a nonempty interval contained in $J$.
    \item $\mathcal H_D=\{(q,L_{q,D},t_D):q\in\partial I_D\}$.
    \item $L_{q,D}$ is a crossing history at
    $q\in\partial I_D$, with every timestamp in $[t_0,t_D)$;
    \item The two histories contain at least one record in
    total, and their latest record is unique. Set
    $t_D=1+\max\{s:(s,x,y,c)\in L_{q,D},\ q\in\partial I_D\}$.
\end{itemize}
We define the initial checkpoint of $\mathcal C$ as
$D_0=(J,\varnothing)$, with $t_{D_0}:=t_0$.
\end{definition}

\begin{definition}[Capped Execution under a Constraint]
\label{def:capped-execution}
We define the \emph{capped-$r$ execution of $M$ on input 
$\mathcal C$ and $z_J$}, which is denoted as $\textbf{Exec}_r(\mathcal C,z_J)$, as follows. 

The input $\mathcal C$ is an execution constraint denoted as
\(
\mathcal C=(J,\sigma_0,i_0,t_0,\mathcal H,\mathcal B,H_L,H_R).
\)
The input $z_J$ is a (random) string over $\{0,1\}^J$. 
The parameter $r$ is an integer with $r\ge1$.

The output is a report  $(\kappa,\sigma,i,t,\gamma)$  defined by \Cref{def:return-report}.  
We emphasize that the execution will stop immediately after returning a report.

We also emphasize here that in our algorithm description, a visit to a position of the random tape refers to the procedure of a maximal consecutive stay at that
    position. The initial placement at an initial position $i_0$ is considered as the
    first visit to $i_0$.
    In a capped-$r$ execution, a position $v\in J$ is
called \emph{exhausted} once its $r$-th visit has ended.


The execution is formally described as follows. It first performs the initialization in \cref{step:init} and 
then runs the repeated transition rule in \cref{step:iteration}
until a report is returned. \cref{step:normal-termination}, \cref{step:new-visit}, and \cref{step:check e against H} define auxiliary
checking rules that are invoked by the first two items.
\begin{enumerate}    
    \item \textit{Initialization.}\label{step:init}
    Set
    \[
    (\sigma,i,t)=(\sigma_0,i_0,t_0).
    \]
    For each $v\in J$, let $N_v$ count the number of visits to $v$.
    Initialize
    \[
    N_{i_0}=1,\qquad
    N_v=0\quad(v\in J\setminus\{i_0\}).
    \]
    Associate history counters $j_q$ ($q\in Q$), $j_L,j_R$
    with $L_q$, $H_L,H_R$, respectively.
    Each history counter gives the index of the next unused
    record in its list. Initialize $j_q=1$ for every $q\in Q$, and set $j_L=j_R=1$.

    If $r\ge2$, set $I=J$ and record the initial checkpoint
    \[
       D_0=(J,\varnothing),\qquad t_{D_0}=t_0.
    \]

    For every $(i_0,\beta)\in\mathcal B$, check
    $z_{i_0}=\beta$.
    If any bit check fails, return $\mathsf{FAIL}$ for the
    failed bit check at $(\sigma_0,i_0,t_0)$ {with $\gamma=(\mathsf{bit},i_0,\beta)$}.
    Otherwise, apply the normal-termination rule defined in \cref{step:normal-termination}.

    \item \label{step:iteration} \textit{Repeated Transition.}
    From $(\sigma,i,t)$, use $z_i$ and the transition rule
    of $M$ to compute $e=(t,i,i',\sigma')$.
    Here $i'$ is the random-head position after the transition,
    and $\sigma'$ is the resulting configuration.

    Check $e$ and $(j_q)_{q \in Q}$ against $\mathcal H$, according to \cref{step:check e against H}.
    If the check fails with report $\gamma$, return
    $\mathsf{FAIL}$ at $(\sigma',i',t+1)$ with $\gamma$.
    Otherwise, apply the normal-termination rule in
    \cref{step:normal-termination} to
    $(\sigma',i',t+1)$. Then:

    \begin{itemize}

    \item If $i'\in J$, perform the following checkpoint
        update when $r\ge2$: if $i\in I$, $N_i=r$, and
        $i'\in I\setminus\{i\}$, write $I=[a,b]$ and set
        \[
           I'=\begin{cases}
               [a,i-1],&i'<i,\\
               [i+1,b],&i'>i.
           \end{cases}
        \]
        Record the checkpoint
        \[
           D=\left(I',\{(q,L_{q,D},t+1):q\in\partial I'\}\right),
           \qquad t_D=t+1,
        \]
        where $L_{q,D}$ is the crossing history at cut $q$
        generated by this execution during $[t_0,t+1)$,
        including the current transition whenever it crosses $q$.
        Set $I\gets I'$.
        In every case, apply \cref{step:new-visit}
        to $(\sigma',i',t+1)$ when $i'\ne i$.
        If no report is returned, set
        \[
            (\sigma,i,t)\gets(\sigma',i',t+1).
        \]

    \item If $i'\notin J$, let $s\in\{L,R\}$ be the exit side to $J$ and
        let $j_s$ be the corresponding counter.
        \begin{itemize}
            \item If $j_s>|H_s|$, return $\mathsf{OVERFLOW}$ at
            $(\sigma',i',t+1)$ {with $\gamma=\bot$}. Otherwise, test $H_s[j_s]=e$.
            If this comparison fails, return $\mathsf{FAIL}$ for
            the failed boundary-history check at
            $(\sigma',i',t+1)$ {with $\gamma=(\mathsf{boundary},s,e)$}. Then set $j_s\gets j_s+1$.

            \item The next record of $H_s$ must have the form
            \[
                e_{\mathrm{in}}=(u,i',i,c),
                 \qquad u\ge t+1.
            \]
            If the next record does not exist or does not have this form, then return $\mathsf{FAIL}$ for the failed boundary-history check at $(\sigma',i',t+1)$, {where we take $\gamma=(\mathsf{boundary},s,H_s[j_s])$ 
            if the record exists, and we take $\gamma=(\mathsf{boundary},s,\bot)$ if otherwise}.

            \item Run \cref{step:check e against H} to check $e_{\mathrm{in}}$ and $(j_q)_{q \in Q}$ against $\mathcal H$. If the check fails
            with report $\gamma$, return $\mathsf{FAIL}$ at
            $(\sigma',i',t+1)$ with $\gamma$.

        \item Set $j_s\gets j_s+1$.
        Set
        \[
           (\sigma,i,t)\gets(c,i,u+1).
        \]
        Apply \cref{step:normal-termination} to $(\sigma,i,t)$.
        Apply \cref{step:new-visit}
        to $(\sigma,i,t)$.
        \end{itemize}

    \item Begin the next iteration by returning to \cref{step:iteration}.

    \end{itemize}

    \item \textit{Normal termination.}\label{step:normal-termination}
    For a given triple $(\sigma,i,t)$, apply the following normal termination rule. Intuitively, this checks whether the execution must stop at $(\sigma,i,t)$.

    The normal termination rule to   $(\sigma,i,t)$ is defined as:
    \begin{itemize}
        \item If $\sigma$ is not halting and $t<T$, continue
        the execution without returning a report.

        \item If $\sigma$ is halting or $t=T$, inspect the
        unused records of all $L_q$, $q\in Q$. If any remain,
        choose a record $e_*$ with the smallest timestamp,
        breaking ties by the smallest cut $q$, and return
        $\mathsf{FAIL}$ at $(\sigma,i,t)$ with
        $\gamma=(\mathsf{history},q,e_*)$.
        If none remain, return $\mathsf{HALT}$ at
        $(\sigma,i,t)$ with $\gamma=\bot$.
    \end{itemize}

    \item \textit{Check a new visit.}\label{step:new-visit}
    This rule checks a new visit inside $J$.
    For a given triple $(\sigma,v,t)$
    with $v\in J$, first check whether $N_v=r$.
    If so, return $\mathsf{OVERFLOW}$ at $(\sigma,v,t)$ {with $\gamma=\bot$}.
    Otherwise, set $N_v\gets N_v+1$ and check
    $z_v=\beta$ for every $(v,\beta)\in\mathcal B$.
    If this check fails, return $\mathsf{FAIL}$ for the
    failed bit check at $(\sigma,v,t)$ {with $\gamma=(\mathsf{bit},v,\beta)$, where $\beta$ is the prescribed value in the failed bit check}.


    \item\textit{Check $e$ and $(j_q)_{q\in Q}$ against $\mathcal H$.} \label{step:check e against H}
Given a record $e=(s,x,y,c)$, indices
$(j_q)_{q\in Q}$ with $1\le j_q\le |L_q|+1$, and $\mathcal{H}$
we define ``check $e$ and $(j_q)_{q\in Q}$ against $\mathcal H$''
as follows.
Consider the cuts $q\in Q$ in increasing order.
If $e$ crosses $q\mid q+1$ with $s<\tau_q$, or if
$j_q\le |L_q|$ and the $j_q$-th record $L_q[j_q]$
has timestamp at most $s$, the check requires that
\[
j_q\le |L_q|
\qquad\text{and}\qquad
L_q[j_q]=e.
\]
If any required comparison fails, the check fails. Let $q$ be the first
failing cut in the prescribed order and set
\[
\gamma=(\mathsf{history},q,e_*),
\qquad
e_*=
\begin{cases}
L_q[j_q],&j_q\le |L_q|,\\
e,&j_q>|L_q|.
\end{cases}
\]
The check succeeds if all required comparisons succeed.
After a successful check, increment $j_q$ for every cut at
which a comparison was required. We call this consuming the
matched records; all other counters remain unchanged.

\end{enumerate}

\end{definition}

\begin{lemma}
\label{lem:initial-execution}
Let $J_0=\{1,2,\ldots,T+1\}$, and let
\[
\mathcal C_0=(J_0,\sigma_{\mathrm{start}},1,0,
              \varnothing,\varnothing,\varnothing,\varnothing)
\]
be the initial constraint, where $\sigma_{\mathrm{start}}$ is the initial configuration of $M$. For every fixed
$z_{J_0}\in\{0,1\}^{J_0}$, the capped-$k$ execution under
$\mathcal C_0$ returns a $\mathsf{HALT}$ report whose
configuration records the output of $M$ on the same random
tape. Consequently, extracting this output from
$\textbf{Exec}_k(\mathcal C_0,U_{J_0})$ gives exactly the
output distribution of $M$.
\end{lemma}

\begin{proof}
Fix $z_{J_0}$. The constraint $\mathcal C_0$ contains no
checks, so the capped execution cannot return
$\mathsf{FAIL}$. Since $M$ visits every random-tape position
at most $k$ times, it never attempts a $(k+1)$-st visit, so
the capped execution cannot overflow inside $J_0$. Moreover,
the head starts at position $1$, moves by at most one cell per
step, and runs for at most $T$ steps. It therefore never
leaves $J_0$, so boundary overflow is also impossible.

Thus the capped execution runs until $M$ halts. At every
step, the two executions read the same random bit and apply
the same transition rule. They consequently have the same
configuration and random-head position throughout, so the
returned configuration records the output of $M$.
Taking uniform $U_{J_0}$ proves the
claim about their output distributions.
\end{proof}

\subsection{The Recursive Sampling Algorithm}
\label{subsec:recursive-sampling}
We now construct $\textbf{Sampler}_r(\mathcal C)$, a
randomized algorithm that uses read-once randomness and
outputs a return report. 
The main purpose of this construction is to let us prove later in \Cref{subsec:sample-correctness} that for every constraint $\mathcal C$ in the constraint families $\mathfrak{C}_r$ defined in \Cref{subsec:sample-correctness}, its
return-report distribution equals that of the corresponding
capped execution on a uniform assignment $U_J$:
\[
 \mathcal L\bigl(\textbf{Sampler}_r(\mathcal C)\bigr)
 =
 \mathcal L\bigl(\textbf{Exec}_r(\mathcal C,U_J)\bigr).
\]

Write
\[
\mathcal C=(J,\sigma_0,i_0,t_0,\mathcal H,
            \mathcal B,H_L,H_R).
\]
At a current checkpoint $D$ under constraint $\mathcal C$,
the sampler guesses the next checkpoint or return report,
together with the information needed to verify it.
We call this tuple a \emph{candidate}.
\begin{definition}[Candidate]
\label{def:candidate}
Given a checkpoint $D=(I_D,\mathcal H_D)$ and an execution constraint $\mathcal C$, writing
$I_D=[a,b]$, a \emph{rank-$r$ candidate} is defined to be a tuple
\[
Y=(\theta,v,d,\beta,\mathcal H_Y)
\]
with the following requirements.
\begin{enumerate}
    \item The field $\theta$ is either the label
    $\mathsf{CONTINUE}$ or a report
    $(\kappa,\sigma,i,t,\gamma)$ in the format of
    \Cref{def:return-report}.

    \item The field $v$ is a position of $I_D$ or $\bot$,
    and $\beta\in\{0,1\}$. If $v=\bot$, then $\beta=0$.
    The field $d$ belongs to $\{\mathsf L,\mathsf R,\bot\}$.

    \item If $\theta=\mathsf{CONTINUE}$, then we require $v\ne\bot$
    and $d\in\{\mathsf L,\mathsf R\}$. Define
    \[
    I'=\begin{cases}
        [a,v-1],&d=\mathsf L,\\
        [v+1,b],&d=\mathsf R,
    \end{cases}
    \]
    and require $I'\ne\varnothing$. 
    Let $S_Y=\partial I'\cup\{v-1,v\}$ be a set of cuts, i.e. it includes the cuts which are at the boundaries of $I'$ and the cuts before and after $v$.
    For every $q\in S_Y$,
    let $L_{q,Y}$ be a crossing history at $q\mid q+1$
    containing at most $2r$ records. Every record has
    timestamp at least $t_0$, and these crossing histories
    contain at least one record in total. Define
    \[
    t_Y=1+\max\{s:(s,x,y,c)\in L_{q,Y},\ q\in S_Y\}
    \]
    and
    $\mathcal H_Y=\{(q,L_{q,Y},t_Y):q\in S_Y\}$.
    By the definition of $t_Y$, some record has timestamp
    $t_Y-1$. Require this record to be unique and to have
    the form $(t_Y-1,v,w,c)$ for some $w\in I'$.

    \item If $\theta$ is a return report
    $(\kappa,\sigma,i,t,\gamma)$, then we require $d=\bot$,
    $\mathcal H_Y=\varnothing$, and $t_Y=t$.
\end{enumerate}
Since $|S_Y| \le 4$, the candidate has an $O(r\mathfrak{b})$-bit
canonical encoding.
\end{definition}

The algorithm proceeds as follows.
Starting from $D_0$, the sampler repeatedly proposes a
candidate for the next checkpoint or return report.
A proposed checkpoint includes the position whose $r$-th
visit ends, its random bit, and the crossing histories
specifying the checkpoint.
To verify the guess, the sampler recursively samples the return
report of a capped-$(r-1)$ execution that replays the computation
through the current checkpoint.  If necessary, the proposed \(r\)-th visit
is then completed and its result is compared with the guess.
If verification fails, the sampler repeats the trial using
fresh independent coins. Otherwise, it updates the current
checkpoint to the one specified by the accepted candidate,
or returns the accepted report and terminates.

\paragraph{\textbf{Sampler}$_r(\mathcal C)$.}
We now give the formal description of the algorithm \textbf{Sampler}$_r(\mathcal C)$. 
The sampler takes an execution constraint $\mathcal C$ and a
visit cap $r\ge1$, and outputs a return report.
It proceeds
as follows, while the subprocedures \textbf{DirectSample}, \textbf{NextGuess}, \textbf{CreateConstraint}, \textbf{FinishAndCheck} will be defined immediately after \textbf{Sampler}.
Write $\mathcal C=(J,\sigma_0,i_0,t_0,\mathcal H, \mathcal B,H_L,H_R)$,
\begin{enumerate}
    \item \label{step:sample-base}
    \emph{Base case.}
    If $r=1$, return $\textbf{DirectSample}(\mathcal C)$.

    \item \label{step:sample-initialize}
    \emph{Initialize the current checkpoint.}
    Let $J$ and $t_0$ be the interval and initial time
    specified by $\mathcal C$.
    Let
    \[
        D_0=(J,\varnothing),\qquad t_{D_0}=t_0,
    \]
    be the initial checkpoint specified by $\mathcal C$.
    Set $D\gets D_0$.

    \item \label{step:sample-trial}
    \emph{Sample a candidate.}
    At the current checkpoint $D$, repeat the following trial
    until it is accepted. In each trial, the proposal and the
    recursive call use independent fresh coins. Whenever a trial
    is rejected, skip its remaining steps and begin a new trial using the same $D$.
    \begin{enumerate}
        \item \label{step:sample-guess}
        \emph{Propose a candidate.}
        Set $Y\gets\textbf{NextGuess}_r(\mathcal C,D)$.
        If $Y=\bot$, reject the trial.

        \item \label{step:sample-create}
        \emph{Prepare the child constraint.}
        Set
        \[
        \mathcal C'\gets\textbf{CreateConstraint}(\mathcal C,D,Y).
        \]
        If $\mathcal C'=\bot$, reject the trial.

        \item \label{step:sample-child}
        \emph{Sample the child report.}
        Set $R\gets\textbf{Sampler}_{r-1}(\mathcal C')$.

        \item \label{step:sample-finish}
        \emph{Verify the candidate.}
        Run
        \[
            \textbf{FinishAndCheck}(\mathcal C,D,Y, \mathcal C', R).
        \]
        Accept the trial if it outputs 1; otherwise reject the trial.
    \end{enumerate}

    \item \label{step:sample-accept}
    \emph{Return a report or continue.}
Let $Y=(\theta,v,d,\beta,\mathcal H_Y)$ be the accepted
candidate (i.e. the candidate corresponding to the accepted trial). 

If $\theta$ is a return report, return $\theta$.
Otherwise, we know $\theta=\mathsf{CONTINUE}$. Denote the 
interval $I_D$ of $D$ as $I_D=[a,b]$ and set
\[
I'=
\begin{cases}
    [a,v-1],&d=\mathsf L,\\
    [v+1,b],&d=\mathsf R.
\end{cases}
\]
Form the new checkpoint by setting
\[
D\gets
\left(
    I',
    \{(q,L_{q,Y},t_Y):q\in\partial I'\}
\right).
\]
Go to \stepref{step:sample-trial} to continue.
\end{enumerate}

We next define the subprocedures.

\paragraph{$\textbf{NextGuess}_r(\mathcal C, D)$.}
Fix a binary encoding space of size
$Q_r=2^{O(r\mathfrak{b})}$, assigning exactly one canonical
encoding to every rank-$r$ candidate for $\mathcal C,D$.
This procedure samples a uniformly random string from this
space. Return the corresponding candidate if the string is
its canonical encoding and satisfies \Cref{def:candidate};
otherwise, return $\bot$.

\paragraph{\textbf{CreateConstraint}$(\mathcal C,D,Y)$.}
On input $(\mathcal C,D ,Y)$, the procedure returns either $\bot$ or a constraint
$\mathcal C'$ on $ I_D$ (which will be used to start the sampling for rank $r-1$). 
Write
\[
 \mathcal C=(J,\sigma_0,i_0,t_0,\mathcal H,\mathcal B,H_L,H_R),
 \qquad D=(I_D,\mathcal H_D),
 \qquad Y=(\theta,v,d,\beta,\mathcal H_Y),
\]
and set $I:=I_D$.
When $D\ne D_0$, write
\[
 \mathcal H_D
 =\{(q,L_{q,D},t_D):q\in\partial I_D\}.
\]
When $\theta$ is a return report, write
$\theta=(\kappa,\sigma,i,t,\gamma)$.
Perform the following steps.
\begin{enumerate}
    \item \label{step:create-compatibility}
    \emph{Check compatibility of $Y$ with $\mathcal C$ and
    $D$.}
    If $\theta=\mathsf{CONTINUE}$, require $t_D<t_Y\le T$.
    Otherwise ($\theta$ is a return report) require
    $t_D\le t_Y\le T$.

    If $\theta$ is a return report and $\kappa = \mathsf{FAIL}$, then
    require $\gamma$ to have one of the following forms,
    with the corresponding condition satisfied:
    \begin{itemize}
        \item $\gamma=(\mathsf{history},q,e)$:
        there is an entry $(q,L_q,\tau_q)\in\mathcal H$.
        \item $\gamma=(\mathsf{bit},u,\alpha)$:
            $(u,\alpha)\in\mathcal B$.
        \item $\gamma=(\mathsf{boundary},s,e)$:
        $s\in\{L,R\}$, referring to the boundary history
        $H_s$ of $\mathcal C$.
    \end{itemize}
    Return $\bot$ if any check fails.

    \item \label{step:create-initial-data}
    \emph{Choose an entrance and bit checks.}
    If $i_0\in I$, set
    $(\sigma'_0,i'_0,t'_0)=(\sigma_0,i_0,t_0)$. Otherwise,
    require the earliest record among the lists $L_{q,D}$
    to be an entrance $e_0=(s,x,y,c)$ with $x\notin I$ and
    $y\in I$, and set
    $(\sigma'_0,i'_0,t'_0)=(c,y,s+1)$. Return $\bot$ if no
    such entrance exists. Let $\mathcal B'$ be the checks of
    $\mathcal B$ whose positions lie in $I$,
    together with $(v,\beta)$ when $v\ne\bot$.

    \item \label{step:create-crossing-requirements}
    \emph{Construct the crossing requirements.}
    Start with the entries of $\mathcal H$.
    For an entry $(q,L_q,\tau_q)$ whose cut lies entirely
    outside $I$, delete from $L_q$ every record with
    timestamp less than $t_D$. Merge the resulting entries
    with all entries of $\mathcal H_D$ and, when
    $\theta=\mathsf{CONTINUE}$, all entries of
    $\mathcal H_Y$.

    The merge retains one entry per cut. To merge
    $(q,L_1,\tau_1)$ and $(q,L_2,\tau_2)$, assume
    $\tau_1\le\tau_2$. Require
    \[
       L_1=L_2\restriction_{<\tau_1},
    \]
    where $L_2\restriction_{<\tau_1}$ consists of the
    records of $L_2$ with timestamp less than $\tau_1$.
    If the equality holds, retain $(q,L_2,\tau_2)$;
    otherwise return $\bot$. Denote the resulting set of
    entries by $\mathcal H'$.

    \item \label{step:create-boundary-histories}
    \emph{Construct the boundary histories.}
    For $s\in\{L,R\}$, set
    \[
    H'_s=
    \begin{cases}
    H_s,&\partial_s I=\partial_s J,\\
    L_{\partial_s I,D},&\partial_s I\ne\partial_s J.
    \end{cases}
    \]

    \item \label{step:create-initial-entrance}
\emph{Check the entrance and prepare the
child's histories.}
If $i_0\in I$, skip this step.
Otherwise, let $e_0=(s,x,y,c)$ be the entrance chosen in
\stepref{step:create-initial-data}, so that
\[
   (\sigma'_0,i'_0,t'_0)=(c,y,s+1).
\]
The child starts immediately after $e_0$ and therefore
will not process this entrance itself. We check $e_0$
against the crossing and boundary histories before
passing their remaining records to the child.

Write
$\mathcal H'=\{(q,L_q,\tau_q):q\in Q'\}$.
Initialize the crossing counters $j_q=1$ for all
$q\in Q'$ and the boundary counters $j_L=j_R=1$.
Perform the following operations.
\begin{enumerate}
    \item \label{step:create-entrance-crossing-check}
    \emph{Apply the crossing check to $e_0$.}
Apply the crossing check in
item~\ref{step:check e against H} of
\Cref{def:capped-execution} to $e_0$,
$(j_q)_{q\in Q'}$, and $\mathcal H'$.
Return $\bot$ if the check fails.

    \item \label{step:create-entrance-boundary-check}
    \emph{Account for $e_0$ in the boundary history.}
    Let $\eta\in\{L,R\}$ be the side through which
    $e_0$ enters $I$.
    As in the supplied-entrance handling in
    item~\ref{step:iteration} of
    \Cref{def:capped-execution}, the entrance record
    must be consumed before the computation starts
    at its resulting configuration.

    Require
    \[
       j_\eta\le |H'_\eta|
       \qquad\text{and}\qquad
       H'_\eta[j_\eta]=e_0.
    \]
    Return $\bot$ if either requirement fails;
    otherwise, increment $j_\eta$.
    Require every remaining record of $H'_L$ and
    $H'_R$ to have timestamp at least $t'_0$.
    Return $\bot$ if this requirement fails.

    \item \label{step:create-entrance-suffix}
    \emph{Retain the records still to be checked.}
    For each $(q,L_q,\tau_q)\in\mathcal H'$, delete
    the first $j_q-1$ records of $L_q$.
    If the resulting list is empty and
    $\tau_q\le t'_0$, remove this entry from
    $\mathcal H'$.
    Retain every other entry with its cutoff unchanged.
    In particular, an empty list with $\tau_q>t'_0$
    still forbids crossings at cut $q$ during
    $[t'_0,\tau_q)$.

    Delete the consumed entrance record from
    $H'_\eta$, leaving the other boundary history
    unchanged.
\end{enumerate}
The child then initializes at
$(\sigma'_0,i'_0,t'_0)$ according to
item~\ref{step:init} of \Cref{def:capped-execution}.
Its initial bit checks and the normal-termination
check are performed during that initialization.

    \item \label{step:create-return}
    \emph{Return the child constraint.}
    Return
    \[
    \mathcal C'
    =(I,\sigma'_0,i'_0,t'_0,
      \mathcal H',\mathcal B',H'_L,H'_R).
    \]
\end{enumerate}

\paragraph{\textbf{FinishAndCheck}$(\mathcal C,D,Y,\mathcal C',R)$.}
For a rank-$r$ call with $r\ge2$, the procedure takes the parent
constraint $\mathcal C$, the current checkpoint $D$, the candidate
$Y$, the child constraint $\mathcal C'\ne\bot$ produced by
$\textbf{CreateConstraint}(\mathcal C,D,Y)$ in the current trial,
and the report $R$ returned by $\textbf{Sampler}_{r-1}(\mathcal C')$.
It returns $1$ if the checks below succeed and $0$ otherwise.
Let $J$ be the interval of $\mathcal C$, and write
\[
 D=(I_D,\mathcal H_D),\qquad I=I_D,\qquad
 Y=(\theta,v,d,\beta,\mathcal H_Y),\qquad
 R=(\kappa_R,\sigma_R,i_R,s,\gamma_R).
\]
A \emph{completion result} is either a return report or a
tuple $(\mathsf{CONTINUE},\sigma',i',t',\bot)$ describing a
nonterminal departure from the proposed $r$-th visit into
$I\setminus\{v\}$.
Perform the following operations to verify that the child
sampling reaches $D$, determine the next outcome by completing
the proposed $r$-th visit when necessary, and check that this
outcome agrees with $Y$.
\begin{enumerate}
    \item \label{step:finish-reach-checkpoint}
    \emph{Verify that the child sampling reaches $D$.}
    If $D=D_0$, skip this step. Otherwise, let
    $(t_D-1,v_D,i_D,\sigma_D)$ be the unique latest record
    in $\mathcal H_D$. Return $0$ if $s<t_D$, since the
    child stopped before the checkpoint was reached.
    If $s>t_D$, proceed to \stepref{step:finish-completion}.
    If $s=t_D$, require the report to be either
    \[
       R=(\mathsf{OVERFLOW},\sigma_D,i_D,t_D,\bot)
    \]
    or
    \[
       R=(\mathsf{FAIL},\sigma_D,i_D,t_D,
          (\mathsf{bit},i_D,\beta_D))
    \]
    for a bit requirement $(i_D,\beta_D)$ of $\mathcal C'$.
    Return $0$ if neither form applies. 

    \item \label{step:finish-completion}
    \emph{Construct a completion result $R'$.}
    The child runs a sampling of level $r-1$, while the parent runs a sampling of level $r$. Thus the child may stop immediately before
    an $r$-th visit that the parent is still allowed to perform.
    In this case, resume from the child's stopping configuration
    and complete that one visit using the bit $\beta$ supplied
    by $Y$. Otherwise, use the child's report directly.
    Denote the resulting outcome by $R'$ and distinguish
    the following two cases.
    \begin{enumerate}
        \item \label{step:finish-no-new-visit}
        \emph{No new $r$-th visit is proposed ($v=\bot$).}
        Return $0$ if $R$ is an $\mathsf{OVERFLOW}$ report
        at a position of $I$; otherwise, set $R'=R$.

        \item \label{step:finish-new-visit}
        \emph{An $r$-th visit to $v$ is proposed ($v\ne\bot$).}
        Perform the following operations to verify that the
        child stopped immediately before this visit, restore
        the history counters, and complete the visit under
        the parent's execution.
        \begin{enumerate}
            \item \label{step:finish-new-visit-report}
            Require
            \[
            R=(\mathsf{OVERFLOW},\sigma_R,v,s,\bot)
            \qquad\text{and}\qquad t_D\le s\le t_Y.
            \]
            Return $0$ if either requirement fails.

            \item \label{step:finish-new-visit-counters}
\label{step:finish-new-visit-execution}
\emph{Complete the proposed $r$-th visit to $v$.}
Let $\mathcal H'$ and $\mathcal B'$ be the crossing
and bit requirements of $\mathcal C'$, respectively,
and let $Q'$ be the set of cuts in $\mathcal H'$.
Whenever a report is produced below, stop this substep
and pass that report, including a $\mathsf{FAIL}$ report,
to \stepref{step:finish-new-visit-result}.

\begin{itemize}
    \item \emph{Restore the history counters.}
    For each crossing or boundary history $L$ of
    $\mathcal C'$, set its next-unused-record counter to
    \[
       1+\bigl|\{e\in L:\operatorname{time}(e)<s\}\bigr|,
    \]
    where $\operatorname{time}(e)$ is the timestamp of $e$.
    Denote the crossing counters by $(j_q)_{q\in Q'}$
    and the boundary counters by $j_L,j_R$.

    \item \emph{Begin the visit.}
    Set
    \[
       (\sigma,i,t)=(\sigma_R,v,s),
       \qquad N_v=r-1.
    \]
    Apply the new-visit check in
    item~\ref{step:new-visit} of
    \Cref{def:capped-execution} to $(\sigma,v,t)$,
    with visit bound $r$, visit counter $N_v$,
    random-tape bit $z_v=\beta$, and bit requirements
    $\mathcal B'$.
    This increments $N_v$ to $r$ and checks
    $\beta=\alpha$ for every $(v,\alpha)\in\mathcal B'$.

    \item \emph{Perform the transitions of this visit.}
    If no report has been produced, repeat the following
    operations.

    From $(\sigma,v,t)$, use $\beta$ and the transition
    rule of $M$ to compute
    \[
       e=(t,v,i',\sigma').
    \]
    Apply the crossing check in
    item~\ref{step:check e against H} of
    \Cref{def:capped-execution} to the record $e$,
    the current counters $(j_q)_{q\in Q'}$,
    and the crossing requirements $\mathcal H'$.
    If this check fails with identifier $\gamma$,
    produce the report
    \[
       (\mathsf{FAIL},\sigma',i',t+1,\gamma).
    \]

    If the crossing check succeeds, apply the
    normal-termination rule in
    item~\ref{step:normal-termination} of
    \Cref{def:capped-execution} to
    $(\sigma',i',t+1)$, using time bound $T$,
    crossing requirements $\mathcal H'$, and the
    updated counters $(j_q)_{q\in Q'}$.

    If no report is produced and $i'=v$, set
    $(\sigma,i,t)=(\sigma',v,t+1)$ and repeat.
    If no report is produced and $i'\ne v$, stop
    this substep and pass the departure configuration
    $(\sigma',i',t+1)$ to
    \stepref{step:finish-new-visit-result}.
    Do not yet apply a new-visit check at $i'$
    or any boundary rule.
\end{itemize}

            \item \label{step:finish-new-visit-result}
            If these checks produce a report, set $R'$ to
            that report. Otherwise, handle the departure
            from $v$. 
            If the head enters
            $i'\in I\setminus\{v\}$ with configuration
            $\sigma'$ at time $t'$, set
            \[
            R'=(\mathsf{CONTINUE},\sigma',i',t',\bot)
            \]
            before the visit to $i'$ begins. 

            If the head enters $i'\in J\setminus I$ with
                configuration $\sigma'$ at time $t'$, set
                \[
                    R'=(\mathsf{OVERFLOW},\sigma',i',t',\bot).
                \]
            
            If the head leaves $J$, apply the
            boundary rules of $\mathcal C$ using the
            corresponding inherited boundary history in
            $\mathcal C'$ and its restored counter, while
            retaining the current crossing checks.
            Set $R'$ to any report produced by these rules,
            including normal termination upon return.
            If they instead resume at $v$ without terminating,
            set $R'$ to the $\mathsf{OVERFLOW}$ report for
            the attempted $(r+1)$-st visit.
        \end{enumerate}
    \end{enumerate}

    \item \label{step:finish-compare}
    \emph{Verify that the outcome agrees with $Y$.}
    Let $t'$ be the time recorded in $R'$. Return $0$ if
    $t'\ne t_Y$. During the completion above, also return
    $0$ if the current simulated time exceeds $t_Y$ before
    a completion result is produced. Check the proposed
    outcome as follows.
    \begin{itemize}
        \item If $\theta=\mathsf{CONTINUE}$, require
        \[
          R'=(\mathsf{CONTINUE},\sigma',i',t_Y,\bot),
          \qquad i'\in I',
        \]
        where $I'$ is the retained interval specified by
        $Y$. Require the unique latest record in
        $\mathcal H_Y$ to be $(t_Y-1,v,i',\sigma')$, and
        require every entry of $\mathcal H_Y$ to have
        been completely checked through time $t_Y$.
        These checks verify both the new checkpoint and
        its proposed crossing histories.

        \item If $\theta$ is a return report, require
        $R'=\theta$, including all fields of the report.
    \end{itemize}
    Return $1$ if all these requirements hold; otherwise,
    return $0$.
\end{enumerate}

\paragraph{\textbf{DirectSample}$(\mathcal C)$.}
Run $\textbf{Exec}_1(\mathcal C,U_J)$, supplying the
random-tape bits on demand.
Whenever the execution first reads a position $v$,
draw one fresh uniform bit and use it for every read
at $v$ during that visit.
Discard the bit when the random head leaves $v$.
Return the report produced by $\textbf{Exec}_1$.

\subsection{Correctness}
\label{subsec:sample-correctness}

We analyze $\textbf{Sampler}_r$ before imposing the time cutoff.
We may assume $k\le T+1$: otherwise, use $T+1$ as the
visit bound throughout the construction, since a computation
with at most $T$ transitions makes at most $T+1$ visits.
The resulting bounds are no larger than those stated
in terms of the original $k$.
Fix the input and the initial constraint $\mathcal C_0$ from
\Cref{lem:initial-execution}. Define families of constraints
$\mathfrak C_k,\ldots,\mathfrak C_1$ recursively as follows.
Set $\mathfrak C_k=\{\mathcal C_0\}$. For $2\le r\le k$,
let $\mathfrak C_{r-1}$ consist of all non-$\bot$ outputs
\[
 \textbf{CreateConstraint}(\mathcal C,D,Y),
 \qquad \mathcal C\in\mathfrak C_r,
\]
where $D$ is a checkpoint reached by
$\textbf{Exec}_r(\mathcal C,z_J)$ for some assignment $z_J$,
including the initial checkpoint, and $Y$ ranges over all rank-$r$ candidates for the
constraint $\mathcal C$ and checkpoint $D$, as defined
in \Cref{def:candidate}.
The candidate may incorrectly describe the next checkpoint
or return report of an execution that reaches $D$.
Thus these families include child constraints constructed
from incorrect guesses. Their
definition uses only the deterministic execution and
constraint-construction procedures, and does not assume
correctness or termination of the sampler.

\begin{lemma}[Boundary histories and counter recovery]
\label{lem:boundary-history-invariant}
For every $1\le r\le k$ and every
\[
\mathcal C=(J,\sigma_0,i_0,t_0,\mathcal H,\mathcal B,H_L,H_R)
\in\mathfrak C_r,
\]
the following statements hold.
\begin{enumerate}
    \item Each nonempty boundary history $H_\eta$ in the
constraint tuple $\mathcal C$, where $\eta\in\{L,R\}$
denotes the left or right boundary, is a prefix of $L_q$
for some entry $(q,L_q,\tau_q)\in\mathcal H$ at the same
boundary cut $q$. Every record of $H_\eta$ has timestamp
less than $\tau_q$.

    \item Suppose a capped-$r$ execution under $\mathcal C$
    reaches a nonterminal checkpoint or returns
    $\mathsf{OVERFLOW}$ at a position inside its interval,
    at time $s$. For every crossing or boundary history
    $L$ of $\mathcal C$, its next-unused-record counter is
    \begin{equation}
      1+\bigl|\{e\in L:\operatorname{time}(e)<s\}\bigr|,
      \label{eq:history-counter-recovery}
    \end{equation}
    where $\operatorname{time}(e)$ denotes the timestamp
    of the record $e$.
\end{enumerate}
\end{lemma}

\begin{proof}
\noindent\emph{First statement: boundary histories.}
We use descending induction on $r$. For $r=k$, the only
constraint is $\mathcal C_0$, whose boundary histories are
empty, so the statement holds.
Fix $2\le r\le k$ and assume the statement holds for every
constraint in $\mathfrak C_r$. Every constraint in
$\mathfrak C_{r-1}$ is a non-$\bot$ output
\[
   \mathcal C'=\textbf{CreateConstraint}(\mathcal C,D,Y)
\]
for some $\mathcal C\in\mathfrak C_r$, reachable checkpoint
$D$, and rank-$r$ candidate $Y$. Fix such a construction and
write $I=I_D$.
All construction steps cited in this part of the proof
are steps of $\textbf{CreateConstraint}$.

First consider the histories after
\stepref{step:create-boundary-histories}, before
\stepref{step:create-initial-entrance} removes any records.
Fix a side $\eta\in\{L,R\}$ for which $H'_\eta$ is
nonempty, and let $q=\partial_\eta I$ be its boundary cut.
The definition of $H'_\eta$ gives two cases.
\begin{itemize}
    \item If $q=\partial_\eta J$, then the child inherits
    this boundary from the parent, and
    \stepref{step:create-boundary-histories} sets
    $H'_\eta=H_\eta$.
    By the induction hypothesis, there is an entry
    $(q,L,\tau)\in\mathcal H$ such that $H_\eta$ is a
    prefix of $L$ and every record of $H_\eta$ has
    timestamp less than $\tau$.
    One of the positions $q,q+1$ belongs to $I$, so this
    cut does not lie entirely outside $I$.
    Thus \stepref{step:create-crossing-requirements}
    includes $(q,L,\tau)$ in the merge without deleting
    any of its records.

    \item If $q\ne\partial_\eta J$, then this is a new
    boundary, and
    \stepref{step:create-boundary-histories} sets
    $H'_\eta=L_{q,D}$.
    The entry $(q,L_{q,D},t_D)$ belongs to $\mathcal H_D$
    and is included in the merge in
    \stepref{step:create-crossing-requirements}.
    Set $L=L_{q,D}$ and $\tau=t_D$. Then $H'_\eta=L$,
    and every record of $H'_\eta$ has timestamp less
    than $\tau$ by \Cref{def:checkpoint}.
\end{itemize}
In either case, the merge receives an entry $(q,L,\tau)$
with the required prefix relation and timestamp bound.
Whenever two entries at $q$ have cutoffs
$\tau_1\le\tau_2$, a successful merge requires
\[
   L_1=L_2\restriction_{<\tau_1}
\]
and retains $(q,L_2,\tau_2)$.
Because the records are ordered by timestamp, $L_1$ is a
prefix of $L_2$. Thus each successful merge preserves both
input lists as prefixes of the retained list and never
decreases either cutoff. Since $\mathcal C'\ne\bot$, all
merges succeed. The resulting entry
$(q,L'_q,\tau'_q)\in\mathcal H'$ therefore has $H'_\eta$
as a prefix of $L'_q$, with every record of $H'_\eta$
having timestamp less than $\tau'_q$.

It remains to check the changes in
\stepref{step:create-initial-entrance}.
If $i_0\in I$, this step is skipped, so the claim already
holds for the returned constraint.
Otherwise, let $e_0$ be the prescribed entrance, let
$\eta_0$ be its entrance side, and let
$q_0=\partial_{\eta_0}I$.
The successful boundary check in
\stepref{step:create-entrance-boundary-check} requires
$H'_{\eta_0}[1]=e_0$.
By the prefix relation proved above,
$L'_{q_0}[1]=e_0$ as well.
Hence \stepref{step:create-entrance-crossing-check} and
\stepref{step:create-entrance-boundary-check} consume the
first record of $L'_{q_0}$ and $H'_{\eta_0}$, respectively.
No other crossing record is consumed: every consumed record
must equal $e_0$, and $e_0$ crosses only cut $q_0$.

Consequently, \stepref{step:create-entrance-suffix} deletes
the same first record from these two lists. Their remaining
suffixes still satisfy the prefix relation. For the other
boundary, neither its history nor the list at its cut loses
any record. All retained cutoffs are unchanged, so the
timestamp bounds also remain valid.
This step can remove an entire crossing entry only when its
remaining list is empty and its cutoff has expired. A
nonempty boundary history cannot be a prefix of an empty
list; thus no entry needed for a nonempty boundary history
is removed. This proves the first statement for $\mathcal C'$
and completes the induction. The argument uses only the
successful construction of $\mathcal C'$, and does not
require $Y$ to describe the actual next outcome.

\medskip
\noindent\emph{Second statement: recovery of the counters.}
Fix $\mathcal C\in\mathfrak C_r$ and a capped-$r$
execution under $\mathcal C$. Every counter starts at $1$
and advances by one whenever its next record is consumed.
It therefore suffices to prove that, at the point specified
in the lemma, the consumed records of each list are exactly
those with timestamp less than $s$.

At the initial checkpoint, $s=t_0$. All stored records have
timestamp at least $t_0$, and no record has been consumed,
so the assertion holds. Now consider any other checkpoint
or an internal overflow at time $s$.
By item~\ref{step:iteration} of
\Cref{def:capped-execution}, the configuration at this
point is reached by a transition or a supplied entrance
with timestamp $s-1$. Its crossing check has succeeded;
otherwise the execution would already have returned
$\mathsf{FAIL}$.

Consider any crossing list $L_q$. Immediately before this
check, its next unused record, if it has timestamp at most
$s-1$, must equal the record being checked, by
item~\ref{step:check e against H} of
\Cref{def:capped-execution}. In particular, its timestamp
must be exactly $s-1$, and the successful check consumes it.
Since timestamps within $L_q$ are strictly increasing,
every record left unused after this check has timestamp
greater than $s-1$. Moreover, no record with timestamp at
least $s$ has been consumed: a crossing record is consumed
only when it equals the transition or supplied entrance
being checked, and all such records processed so far have
timestamp at most $s-1$. Thus the consumed records of
$L_q$ are exactly those with timestamp less than $s$.

Next consider a nonempty boundary history $H_\eta$ at cut
$q=\partial_\eta J$. By the first statement, it is a
prefix of a crossing list $L_q$. Every record of $H_\eta$
with timestamp less than $s$ has therefore been matched
by a successful crossing check. Such a record describes
either an exit from $J$ or a supplied entrance into $J$.
In the exit case, item~\ref{step:iteration} of
\Cref{def:capped-execution} then compares the same record
with the next unused record of $H_\eta$ and consumes it.
If this boundary check failed, or if the required record
were absent, the execution would stop outside $J$ before
reaching the checkpoint or internal overflow under
consideration. In the entrance case, the boundary record
is consumed after its successful crossing check and before
its resulting configuration is installed.
Hence every boundary record with timestamp less than $s$
has also been consumed in its boundary history.

It remains to show that no boundary record with timestamp
at least $s$ has been consumed.
Let $e$ be any boundary record already consumed at the
checkpoint or internal overflow under consideration,
and let $u$ be its timestamp.
If $e$ is an exit record, it is consumed after the
transition from time $u$ to time $u+1$.
If $e$ is a supplied entrance record, it is consumed
immediately before the configuration at time $u+1$ is
installed; no checkpoint or internal overflow is produced
between these two operations.
In either case, since time never decreases, the checkpoint
or internal overflow under consideration must satisfy
$s\ge u+1$. Hence $u<s$.

Thus the consumed boundary records are exactly those with
timestamp less than $s$. An empty boundary history has
counter $1$ throughout. Together with the argument for
crossing histories, this proves
\eqref{eq:history-counter-recovery}.
\end{proof}

For the next three lemmas, fix
$2\le r\le k$ and $\mathcal C\in\mathfrak C_r$ on $J$.
For the analysis only, number the stages by $j$.
Stage $0$ starts at the initial checkpoint
$D_0=(J,\varnothing)$, and each accepted
$\mathsf{CONTINUE}$ candidate advances to the next stage.
Rejected trials do not advance $j$.
For $\textbf{Exec}_r$, each recorded checkpoint after
$D_0$ likewise advances the stage by one.

Drawing the random assignment $U_J$ induces a random checkpoint
path (a sequence of checkpoints). Let $D_j$ be its checkpoint at stage $j$, with
$D_0=(J,\varnothing)$, and set $D_j=\bot$ if stage $j$ is not
reached. On the event $D_j\ne\bot$, let $O_j$ be the next
outcome: $D_{j+1}$ if the execution continues, and its return
report if it stops. Fix a stage $j$ and a checkpoint
$D$ reachable at that stage, and define
\[
 A_{j,D}=\{z_J\in\{0,1\}^J:
 \textbf{Exec}_r(\mathcal C,z_J)\text{ reaches $D$ at stage $j$}\},
 \qquad I=I_D.
\]
Thus $U_J\in A_{j,D}$ is the event that the capped execution
reaches $D$ at stage $j$. For every $I\subseteq J$, write
$U_I:=U_J|_I$.
We write $\mathcal L(U_I\mid A_{j,D})$ for the conditional
distribution of $U_I$ given the event $U_J\in A_{j,D}$.
Equivalently, choose an assignment $z_J$ uniformly from
$A_{j,D}$ and return its restriction $z_J|_I$.
Since $D$ is reachable at stage $j$, the set $A_{j,D}$ is
nonempty. Thus, for every $u\in\{0,1\}^I$,
\[
 \Pr[U_I=u\mid U_J\in A_{j,D}]
 =\frac{\bigl|\{z_J\in A_{j,D}:z_J|_I=u\}\bigr|}
        {|A_{j,D}|}.
\]

We first determine the conditional distribution of the random
bits in the current interval after the execution reaches $D$.

\begin{lemma}
\label{lem:boundary-factorization}
There are sets $E_D\subseteq\{0,1\}^I$ and
$F_D\subseteq\{0,1\}^{J\setminus I}$ such that
\begin{equation}
 A_{j,D}
 =\{z_J\in\{0,1\}^J:
       z_J|_I\in E_D,\ z_J|_{J\setminus I}\in F_D\}.
 \label{eq:boundary-factorization}
\end{equation}
Consequently, with $q_D=\Pr[U_I\in E_D]>0$, we have
$\Pr[U_J\in A_{j,D}]\le q_D$ and
\begin{equation}
 \mathcal L(U_I\mid A_{j,D})=\operatorname{Unif}(E_D).
 \label{eq:conditional-interior-law}
\end{equation}
\end{lemma}

\begin{proof}
We proceed in three steps.
\begin{enumerate}
    \item \emph{Define the two sets.}
    Let
    \[
       E_D=\{z_J|_I:z_J\in A_{j,D}\},\qquad
       F_D=\{z_J|_{J\setminus I}:z_J\in A_{j,D}\}.
    \]
    Since $D$ is reachable at stage $j$, the set $A_{j,D}$
    is nonempty, and hence so are $E_D$ and $F_D$.
    By definition, every assignment in $A_{j,D}$ restricts
    to an element of $E_D$ on $I$ and an element of $F_D$
    on $J\setminus I$. This proves the left-to-right
    inclusion in \eqref{eq:boundary-factorization}.

    \item \emph{Prove the reverse inclusion.} \label{item:2}
    If $D=D_0$, then $j=0$, $I=J$, and
    $A_{0,D_0}=\{0,1\}^J$, so the reverse inclusion is
    immediate. Suppose $D\ne D_0$.
    Fix $u\in E_D$ and $u'\in F_D$. Choose
    $z,\widetilde z\in A_{j,D}$ such that $z|_I=u$ and
    $\widetilde z|_{J\setminus I}=u'$, and define $w$ by
    \[
       w|_I=z|_I=u,\qquad
       w|_{J\setminus I}=\widetilde z|_{J\setminus I}=u'.
    \]
    We show that the capped-$r$ execution on $w$ reaches
    $D$ at stage $j$.

    Both executions on $z$ and $\widetilde z$ reach
    $D=(I,\mathcal H_D)$. For each $q\in\partial I$,
    the checkpoint therefore records the same complete
    crossing history $L_{q,D}$ during $[t_0,t_D)$.
    In particular, a boundary crossing with record
    $(s,x,y,c)$ occurs in both executions and leaves both
    at the same triple $(c,y,s+1)$.
    Starting from the common initial data of $\mathcal C$,
    follow the execution on $z$ inside $I$ and the execution
    on $\widetilde z$ outside $I$.
    Between boundary crossings, the bits read agree with
    $w$. At an exit from $I$, perform the departure from
    the execution on $z$; its resulting triple agrees
    with that of $\widetilde z$, so continue with the
    latter execution outside $I$. At an entrance into $I$,
    follow the entrance from $\widetilde z$; its resulting
    triple agrees with that of $z$, so continue with the
    latter execution inside $I$.
    Excursions outside $J$ use the fixed boundary histories
    of $\mathcal C$. Those crossing a boundary of $I$
    have the common exit and entrance records just
    described; all others belong to the unchanged exterior
    execution on $\widetilde z$.

    We check that this combined computation obeys the
    execution rules. At each position in $I$, its visits
    and random bit are those of $z$; at each position in
    $J\setminus I$, they are those of $\widetilde z$.
    At a cut entirely inside $I$, its crossing history
    is that of $z$; at a cut entirely outside $I$, it is
    that of $\widetilde z$; at a boundary of $I$, it is
    the common history in $\mathcal H_D$.
    Thus the visit and bit checks, crossing checks, and
    boundary rules needed to reach the checkpoint update
    at time $t_D$ all succeed, as in the corresponding
    original executions. No earlier normal termination
    occurs, since each configuration and its timestamp
    come from an original execution before it reaches $D$.
    The combined computation is therefore the capped-$r$
    execution under $\mathcal C$ on $w$ up to this update.

    To identify its checkpoint, first observe that no
    position of $I$ completes its $r$-th visit before $D$
    in either original execution. Such a departure would
    either produce a checkpoint excluding that position
    from every later interval, or cause a return report
    before another checkpoint could be recorded. Both
    possibilities contradict reaching $D$ with that
    position still in $I$.
    All departures completing an $r$-th visit through
    time $t_D$ consequently occur in $J\setminus I$.
    In the combined computation, these departures have
    exactly the same times, positions, and directions as
    in the execution on $\widetilde z$.
    The checkpoint update rule therefore produces the same
    sequence of retained intervals and reaches $I$ at
    stage $j$ and time $t_D$. Its crossing histories at
    the boundaries of $I$ are $\mathcal H_D$, so its
    checkpoint at that stage is exactly $D$.
    Hence $w\in A_{j,D}$. Since $u$ and $u'$ were
    arbitrary, this proves the reverse inclusion and
    establishes \eqref{eq:boundary-factorization}.

    \item \emph{Compute the probabilities.}
    By \eqref{eq:boundary-factorization}, pairing any
    assignment in $E_D$ with any assignment in $F_D$
    gives an assignment in $A_{j,D}$, and every assignment
    in $A_{j,D}$ is obtained uniquely in this way. Thus
    $|A_{j,D}|=|E_D|\,|F_D|$. Since $U_J$ is uniform
    and $q_D=|E_D|/2^{|I|}$,
    \[
       \Pr[U_J\in A_{j,D}]
       =\frac{|E_D|\,|F_D|}{2^{|J|}}
       =q_D\frac{|F_D|}{2^{|J\setminus I|}}
       \le q_D.
    \]
    For each fixed $u\in E_D$, there are exactly $|F_D|$
    assignments in $A_{j,D}$ whose restriction to $I$
    is $u$, one for each assignment in $F_D$. Therefore
    \[
       \Pr[U_I=u\mid U_J\in A_{j,D}]
       =\frac{|F_D|}{|A_{j,D}|}
       =\frac{1}{|E_D|}.
    \]
    For $u\notin E_D$, the conditional probability is
    $0$. Hence
    $\mathcal L(U_I\mid A_{j,D})=\operatorname{Unif}(E_D)$,
    proving \eqref{eq:conditional-interior-law}.
\end{enumerate}
\end{proof}

We next specify how to form a candidate from an execution
that reaches $D$, and prove that the resulting candidate
depends only on the random bits in $I_D$.

\begin{lemma}[Determination of the canonical candidate]
\label{lem:canonical-candidate}
Fix $2\le r\le k$, a constraint $\mathcal C\in\mathfrak C_r$,
and a checkpoint $D$ reachable at stage $j$. Write $I=I_D$.
Fix $z_I\in E_D$. For each extension $z_J\in A_{j,D}$
with $z_J|_I=z_I$, follow
$\textbf{Exec}_r(\mathcal C,z_J)$ through $D$ to its next
checkpoint or return report, and form the rank-$r$
candidate specified in
item~\ref{step:canonical-candidate-fields} of the proof.
This candidate is independent of the chosen extension.
We call it the \emph{canonical candidate} for
$\mathcal C,D,z_I$.
\end{lemma}

\begin{proof}
By \Cref{lem:boundary-factorization}, every $z_I\in E_D$
has an extension in $A_{j,D}$. Fix any two such extensions
$z,\widetilde z$. We first identify the portions of their
executions that agree, and then specify the candidate fields.
\begin{enumerate}
    \item \emph{Determine the execution inside $I$ through $D$.}
    If $i_0\in I$, both executions start with the initial
    data of $\mathcal C$ and read the same bits $z_I$.
    Otherwise, both first enter $I$ according to the same
    earliest entrance record in $\mathcal H_D$.
    Each subsequent entrance recorded in $\mathcal H_D$
    also places both executions at the same configuration,
    position, and time. Between entrances and exits, the
    transitions inside $I$ are determined by these data
    and $z_I$. Hence the executions have the same visits
    to every position of $I$ and the same crossing histories
    at every cut inside $I$ through $D$.
    Their configuration, head position, and time at $D$
    agree, and their visit counters in $I$ agree at that
    point. By \Cref{lem:boundary-history-invariant}, all
    crossing and boundary counters of $\mathcal C$ also
    agree, since they are determined by the stored lists
    and $t_D$.

    \item \emph{Determine the next outcome after $D$.}
    Resume the execution from $D$, including any checks
    still pending at that checkpoint. While the head stays
    in $I$, both executions use the same bits, configurations,
    visit counters, and history counters, so they perform
    the same transitions and checks.
    Any entry into $J\setminus I$ must first reach a position
$v$ adjacent to $I$. This position has already completed
its $r$-th visit, so $N_v=r$. Unless the preceding crossing
checks or normal-termination rule have already returned
a report, the new-visit check in
item~\ref{step:new-visit} of \Cref{def:capped-execution}
returns $\mathsf{OVERFLOW}$ before reading $z_v$.
Thus the execution reads no random bit in $J\setminus I$
after reaching $D$.
    An excursion outside $J$ uses the fixed boundary
    histories of $\mathcal C$ with the same counters in
    both executions. Consequently, the two executions
    agree until they reach the next checkpoint or produce
    a return report, including all fields of that report.

    \item \label{step:canonical-candidate-fields}
    \emph{Specify every field of the candidate.}
    Set $\theta=\mathsf{CONTINUE}$ if the next outcome is
    a checkpoint, and otherwise let $\theta$ be the return
    report. If an $r$-th visit to a position of $I$ begins
    at or after $D$ before this outcome is produced, let
    $v$ be that position and set $\beta=(z_I)_v$; otherwise,
    set $v=\bot$ and $\beta=0$.
    There is at most one such visit: its departure either
    records the next checkpoint or leads to a return report
    before another $r$-th visit in $I$ can begin.
    These fields agree by the preceding step.
    If $\theta$ is a return report, set $d=\bot$ and
    $\mathcal H_Y=\varnothing$, as in \Cref{def:candidate}.
    Otherwise, let $d$ be the actual departure direction
    from $v$, and let $\mathcal H_Y$ contain the actual
    crossing histories at the cuts in $S_Y$ during
    $[t_0,t_Y)$, where $t_Y$ is the new checkpoint time.
    Every cut in $S_Y$ lies inside $I$
    or at a boundary of $I$. Its history before $D$ is
    determined by the first step and $\mathcal H_D$, and
    its subsequent records are determined by the second
    step. Thus these histories also agree.
    The resulting tuple satisfies \Cref{def:candidate};
    in particular, each relevant cut has at most $2r$
    crossings before this nonterminal outcome.
\end{enumerate}
Therefore all fields of the canonical candidate are
independent of the chosen extension of $z_I$.
\end{proof}

For $z_I\in E_D$, write $g_D(z_I)$ for the canonical
candidate specified by \Cref{lem:canonical-candidate},
with $\mathcal C$ understood from the context. Set
$g_D(z_I)=\bot$ when $z_I\notin E_D$.
For a candidate $Y=(\theta,v,d,\beta,\mathcal H_Y)$, let
$\pi(Y)$ be the next checkpoint specified by $Y$ when
$\theta=\mathsf{CONTINUE}$, and the return report $\theta$
otherwise. Set $\pi(\bot)=\bot$.

We next analyze $\textbf{CreateConstraint}$ and
$\textbf{FinishAndCheck}$ for a fixed assignment $z_I$.
For a candidate $Y$, write
$\mathcal C'=\textbf{CreateConstraint}(\mathcal C,D,Y)$.
The following lemma proves that $\mathcal C'\ne\bot$
whenever $Y=g_D(z_I)$. Whenever $\mathcal C'\ne\bot$,
it also proves that $\textbf{FinishAndCheck}$, given the
child report $\textbf{Exec}_{r-1}(\mathcal C',z_I)$,
returns $1$ if and only if $Y=g_D(z_I)$.

\begin{lemma}[Exact verification of a candidate]
\label{lem:outcome-test}
Fix $2\le r\le k$, a constraint $\mathcal C\in\mathfrak C_r$,
a checkpoint $D$ reachable at stage $j$, and a rank-$r$
candidate $Y$. Write $I=I_D$ and let
\[
   \mathcal C'=\textbf{CreateConstraint}(\mathcal C,D,Y).
\]
If $g_D(z_I)=Y$ for some $z_I\in E_D$, then
$\mathcal C'\ne\bot$. Whenever $\mathcal C'\ne\bot$, we have
$\mathcal C'\in\mathfrak C_{r-1}$ and, for every
$z_I\in\{0,1\}^I$,
\begin{equation}
 \textbf{FinishAndCheck}\bigl(\mathcal C,D,Y,\mathcal C',
       \textbf{Exec}_{r-1}(\mathcal C',z_I)\bigr)=1
 \quad\Longleftrightarrow\quad
 g_D(z_I)=Y.
 \label{eq:complete-test-characterization}
\end{equation}
\end{lemma}

\begin{proof}
We first prove that a canonical candidate gives
$\mathcal C'\ne\bot$, and that every non-$\bot$ child
constraint belongs to $\mathfrak C_{r-1}$.
We then prove the two implications in
\eqref{eq:complete-test-characterization} separately.
\begin{enumerate}
    \item \label{step:verification-construction}
    \emph{The construction succeeds for a canonical candidate.}
    Suppose $g_D(z_I)=Y$. By \Cref{lem:canonical-candidate},
    choose an extension $w\in A_{j,D}$ with $w|_I=z_I$.
    The candidate formed from this execution is $Y$.
    We verify the checks in
    $\textbf{CreateConstraint}$, as defined in
    \Cref{subsec:recursive-sampling}, using the capped-$r$
    execution $\textbf{Exec}_r(\mathcal C,w)$.
    \begin{enumerate}
        \item \emph{Compatibility with $\mathcal C$ and $D$.}
        In \stepref{step:create-compatibility} of
        $\textbf{CreateConstraint}$, the time inequalities
        hold because $Y$ records the next outcome after
        $D$: a subsequent checkpoint has $t_Y>t_D$, and
        a return report has $t_Y\ge t_D$. Every reported
        time is at most $T$ by \Cref{def:capped-execution}.
        Since \(Y=g_D(z_I)\) is the canonical candidate arising from \(\mathbf{Exec}_r(\mathcal C,w)\), if its field \(\theta\) has tag \(\mathsf{FAIL}\), then \(\theta\) is the return report produced by that execution.
        By \Cref{def:return-report}, its identifier
        $\gamma$ identifies an entry
        $(q,L_q,\tau_q)\in\mathcal H$, a bit requirement
        $(u,\alpha)\in\mathcal B$, or a boundary history
        $H_\eta$ with $\eta\in\{L,R\}$.
        Thus $\gamma$ satisfies the corresponding condition
        in this step.

        \item \emph{The initial data and bit requirements.}
        Consider \stepref{step:create-initial-data} of
        $\textbf{CreateConstraint}$.
        If \(i_0\in I\), CreateConstraint sets the child’s initial configuration, head position, and time equal to those of the parent constraint \(\mathcal C\).
        Otherwise, since the execution reaches $D$ with its
        head in $I$, it must first enter $I$.
        Its earliest boundary record in $\mathcal H_D$
        is precisely that entrance $e_0=(s,x,y,c)$.
        Thus the required entrance exists and specifies
        $(\sigma'_0,i'_0,t'_0)=(c,y,s+1)$.
        When $v\ne\bot$, the additional bit requirement
        is $(v,w_v)$, since $Y$ is canonical.
        Forming $\mathcal B'$ involves no further check.

        \item \emph{Compatibility of the crossing histories.}
        Consider \stepref{step:create-crossing-requirements}
        of $\textbf{CreateConstraint}$.
        Every cut in $\mathcal H_D$ or $\mathcal H_Y$ has
        an endpoint in $I$. Hence the deletion of records
        at cuts entirely outside $I$ does not change any
        entry that will be merged with these histories.
        For each \(q\in\partial I_D\), the list \(L_{q,D}\) consists of exactly the crossing records generated by \(\mathbf{Exec}_r(\mathcal C,w)\) at cut \(q\mid q+1\) during \([t_0,t_D)\), in chronological order.
        When \(\theta=\mathsf{CONTINUE}\), for each \(q\in S_Y\), the list \(L_{q,Y}\) consists of exactly the crossing records generated by \(\mathbf{Exec}_r(\mathcal C,w)\) at cut \(q\mid q+1\) during \([t_0,t_Y)\), in chronological order.
        By the crossing check in
        item~\ref{step:check e against H} of
        \Cref{def:capped-execution}, the inherited
        requirements agree with these actual histories
        throughout their common time intervals: For each inherited requirement \((q,L_q,\tau_q)\), before a checkpoint at time \(t_*\) is reached, the crossing checks ensure that the records in \(L_q\) with timestamps less than \(\min\{\tau_q,t_*\}\) are exactly the crossing records generated by the execution at cut \(q\mid q+1\) during \([t_0,\min\{\tau_q,t_*\})\).
       Therefore, whenever two entries \((q,L_1,\tau_1)\) and \((q,L_2,\tau_2)\) refer to the same cut and \(\tau_1\le\tau_2\), the list \(L_1\) consists of exactly those records in \(L_2\) whose timestamps are less than \(\tau_1\), in the same order. Thus,
        \[
            L_1=L_2\restriction_{<\tau_1}.
        \]
        This equality is precisely the condition required to merge the two entries, after which the procedure keeps \((q,L_2,\tau_2)\). The same argument shows that this retained entry agrees with each remaining entry at the same cut on their common time interval, so every subsequent merge also succeeds.

        \item \emph{The entrance checks and remaining histories.}
        The boundary histories are assigned in
        \stepref{step:create-boundary-histories} of
        $\textbf{CreateConstraint}$ without any rejection.
        If $i_0\in I$,
        \stepref{step:create-initial-entrance} is skipped.
        Suppose $i_0\notin I$, and use the entrance $e_0$
        from the preceding argument. Let $q_0$ be its cut,
        let $\eta$ be its entrance side, and write
        $\mathcal H'=\{(q,L'_q,\tau'_q):q\in Q'\}$.
        Before this first entrance, the random-tape head remains outside \(I\). Thus, no earlier transition crosses either boundary of \(I\) or any cut between two positions in \(I\).
        At cuts entirely
        outside $I$, all records with timestamp less than
        $t_D$ have already been deleted. Since
        $t'_0=s+1\le t_D$, the merged lists satisfy
        \[
           L'_q\restriction_{<t'_0}
           =\begin{cases}
               (e_0),&q=q_0,\\
               (),&q\ne q_0.
             \end{cases}
        \]
        Moreover, the entry from $\mathcal H_D$ at $q_0$
        ensures $\tau'_{q_0}\ge t_D>s$.
        Thus the crossing check in
        item~\ref{step:check e against H} of
        \Cref{def:capped-execution}, applied to $e_0$
        with all counters initialized to $1$, matches
        exactly the first record of $L'_{q_0}$.
        This proves that
        \stepref{step:create-entrance-crossing-check} of
        $\textbf{CreateConstraint}$ succeeds.

        Since $i_0\in J\setminus I$ and every supplied
        excursion outside $J$ returns to the endpoint from
        which it exited, as specified in
        item~\ref{step:iteration} of
        \Cref{def:capped-execution}, the first entrance
        $e_0$ crosses a new boundary of $I$.
        Thus $H'_\eta=L_{q_0,D}$ begins with $e_0$.
        Any other newly assigned boundary history is also
        taken from $\mathcal H_D$ and has no record before
        $t'_0$. For an inherited boundary history,
        \Cref{lem:boundary-history-invariant} implies that
        all records with timestamp less than $t_D$ have
        been consumed in the parent execution. Since no
        crossing of that boundary occurs before the first
        entrance into $I$, it likewise has no record before
        $t'_0$. Hence, after $e_0$ is removed, every remaining
        boundary record has timestamp at least $t'_0$.
        These are exactly the requirements in
        \stepref{step:create-entrance-boundary-check} of
        $\textbf{CreateConstraint}$.
        Finally, \stepref{step:create-entrance-suffix} of
        that procedure only deletes consumed records and
        expired empty entries, so it cannot reject.
    \end{enumerate}
    All checks in $\textbf{CreateConstraint}$ therefore
    succeed, and \stepref{step:create-return} of that
    procedure returns $\mathcal C'\ne\bot$.
    This proves the first assertion of the lemma.
    Independently of whether $Y$ is canonical, whenever
    $\mathcal C'\ne\bot$, the definition of the constraint
    families gives $\mathcal C'\in\mathfrak C_{r-1}$.
    For the remaining two parts, fix a candidate with
    $\mathcal C'\ne\bot$ and an arbitrary
    $z_I\in\{0,1\}^I$, and write
    \[
       R=\textbf{Exec}_{r-1}(\mathcal C',z_I).
    \]

    \item \emph{Acceptance implies $Y=g_D(z_I)$.}
    Assume
    \[
       \textbf{FinishAndCheck}(\mathcal C,D,Y,\mathcal C',R)=1.
    \]
    We first show that $z_I\in E_D$, and then identify
    every field of the accepted candidate.
    \begin{enumerate}
        \item \label{step:verification-reach}
        \emph{The parent execution reaches $D$.}
        Fix $w^*\in A_{j,D}$ and define $w$ by
        \[
           w|_I=z_I,\qquad
           w|_{J\setminus I}=w^*|_{J\setminus I}.
        \]
        If $D=D_0$, then $I=J$ and $w\in A_{0,D_0}$
        follows immediately. Suppose $D\ne D_0$.
        By \stepref{step:create-initial-data} of
        $\textbf{CreateConstraint}$, the child starts with
        the parent's initial data when $i_0\in I$, and
        otherwise immediately after the first prescribed
        entrance into $I$. In the latter case,
        \stepref{step:create-initial-entrance} of that
        procedure checks and consumes this entrance.
        The remaining records of $\mathcal H_D$ are
        prescribed by the child's crossing requirements,
        and its boundary histories supply the prescribed
        exits and entrances through time $t_D$.

        Since the test accepts,
        \stepref{step:finish-reach-checkpoint} of
        $\textbf{FinishAndCheck}$ succeeds. In particular,
        the child cannot return before $t_D$, so it does
        not attempt an $r$-th visit inside $I$ before
        that time. The check in
        item~\ref{step:check e against H} of
        \Cref{def:capped-execution} requires each record
        of $\mathcal H_D$ to be matched at its timestamp.
        A transition inconsistent with its last record
        gives a history failure at time $t_D$, which the
        reach-checkpoint step rejects. A supplied entrance
        is checked before its resulting configuration is
        installed. If that entrance skips a required record,
        the failure is reported at the time immediately
        after the preceding exit, so it cannot certify
        that the skipped checkpoint was reached.
        If the child stops exactly at $t_D$, the permitted
        report forms specify the configuration and position
        of $D$ and allow only an internal overflow or a
        bit failure at that position.

        Hence the child's execution inside $I$ joins the
        execution on $w^*$ outside $I$ at the records of
        $\mathcal H_D$. As in the proof of
        \Cref{lem:boundary-factorization}, these portions
        form the parent execution on $w$ through $D$ at
        stage $j$: all earlier departures ending an
        $r$-th visit occur outside $I$ and agree with
        those of $w^*$.
        Thus $w\in A_{j,D}$ and $z_I\in E_D$.
        By \Cref{lem:canonical-candidate}, the next
        canonical candidate of this parent execution is
        $g_D(z_I)$.

\item \label{step:verification-transition-comparison}
\emph{Identify the report returned by the child execution.}

Recall that
\[
R=\textbf{Exec}_{r-1}(\mathcal C',z_I).
\]
Under the assumption that $\textbf{FinishAndCheck}$ accepts,
we prove the following 3 statements.
\begin{enumerate}
    \item If
    $R=(\mathsf{OVERFLOW},\sigma,p,s,\bot)$ with $p\in I$,
    then $\textbf{Exec}_r(\mathcal C,w)$ reaches
    configuration $\sigma$ and position $p$ at time $s$,
    immediately before its first new $r$-th visit in $I$
    at or after $D$.
    The child stops because its visit bound is $r-1$,
    whereas the parent is allowed to perform this visit.

    \item Otherwise, $R$ is exactly the next report
    returned by $\textbf{Exec}_r(\mathcal C,w)$ at or
    after $D$, before any new $r$-th visit in $I$ begins.

    \item During the completion (\stepref{step:finish-new-visit}) of
$\textbf{FinishAndCheck}$, every check that comes only
from the requirements added by $D$ or $Y$ succeeds.
\end{enumerate}

By the preceding item,
$\textbf{Exec}_r(\mathcal C,w)$ and
$\textbf{Exec}_{r-1}(\mathcal C',z_I)$ reach the
configuration and random-head position specified by $D$
at time $t_D$.
For each position in $I$, their visits before $t_D$
occur at the same times.
If the child starts immediately after the first entrance
into $I$, that entrance has already been checked in
\stepref{step:create-initial-entrance} of
$\textbf{CreateConstraint}$.

We first examine the crossing requirements in $\mathcal C'$.
At cuts entirely outside $I$,
\stepref{step:create-crossing-requirements} deletes
records with timestamps less than $t_D$.
The parent has already checked these records before
reaching $D$.
At the other cuts, this step deletes no records.
When an entry $(q,L_q,\tau_q)$ is merged with another
entry, the retained list agrees with $L_q$ at timestamps
less than $\tau_q$, after the stated deletions.
The retained entry also has an ending time at least
$\tau_q$.
Thus, at timestamps at least $t_D$, the child still
requires every crossing prescribed by the inherited
entry and forbids every unlisted crossing before $\tau_q$.
The entrance step removes only records that it has
already checked.

The records added from $\mathcal H_D$ have timestamps
less than $t_D$ and have already been checked.
The corresponding requirements end at $t_D$.
They therefore place no restriction on crossings
with timestamps at least $t_D$.

\emph{Proof of statement \textup{i}.}
Suppose
\[
R=(\mathsf{OVERFLOW},\sigma,p,s,\bot),
\qquad p\in I.
\]
Since $\textbf{FinishAndCheck}$ accepts,
\stepref{step:finish-no-new-visit} implies $v\ne\bot$,
and \stepref{step:finish-new-visit-report} gives
$v=p$ and $t_D\le s\le t_Y$.

The child returns $R$, so it has not returned any
earlier report.
In particular, every added check that the child performs
before returning $R$ succeeds.
We show that the parent reaches configuration $\sigma$
and position $p$ at time $s$, immediately before its
first new $r$-th visit in $I$ at or after $D$.

Suppose both executions are in configuration $c$,
with their heads at $u\in I$, at time $t$, and are
about to perform a transition.
Since $w_u=(z_I)_u$, both compute the same record
\[
e=(t,u,u',c').
\]
The child checks all crossing requirements of the
parent that remain to be checked, together with any
additional requirements.
Since the child's crossing check succeeds, the parent's
crossing check also succeeds.
Neither execution can stop by normal termination here:
a halting configuration or time $T$ would make the
child return $\mathsf{HALT}$ or $\mathsf{FAIL}$,
rather than $R$.

We then apply the new-visit or boundary rules.
\begin{itemize}
    \item If $u'=u$, neither execution begins a new visit.
    Both continue from $(c',u,t+1)$.

    If $u'\in I\setminus\{u\}$, the numbers of previous
    visits to $u'$ are equal in both executions.
    If this number is less than $r-1$, both allow
    the new visit.
    The child checks all bit requirements of the parent
    at $u'$, as well as any added bit requirement.
    These checks succeed because otherwise the child
    would return $\mathsf{FAIL}$.
    Thus both executions continue from $(c',u',t+1)$.

    If the number of previous visits is $r-1$, the
    child returns
    \[
    (\mathsf{OVERFLOW},c',u',t+1,\bot)
    \]
    before checking the bit at $u'$.
    The parent reaches the same configuration and
    position at time $t+1$, before its new-visit check,
    and is allowed to begin its $r$-th visit there.

    \item If $u'\in J\setminus I$, the child's boundary
    history at this boundary is taken from $D$.
    All its records have already been checked.
    Hence the child returns $\mathsf{OVERFLOW}$ at
    $u'\notin I$, contrary to the assumed form of $R$.
    In the parent execution, $u'$ has already completed
    its $r$-th visit, so the parent would return the
    same overflow report.

    \item If $u'\notin J$, the boundary crossed is a
    boundary of both $I$ and $J$.
    Both executions therefore use the same unused
    records of the boundary history inherited from
    $\mathcal C$.
    They inspect the same exit and entrance records.
    Every crossing requirement of the parent that
    must be checked against the entrance is also
    imposed on the child.
    A failed boundary or crossing check would stop
    the child before it returns $R$.
    Thus both accept the same entrance and set their
    configuration, position, and time to those specified
    by that entrance.
    They then apply normal termination and the
    new-visit rule described above.
\end{itemize}

The same reasoning applies to any initialization or
new-visit checks still pending at time $t_D$.
Repeating these arguments shows that the parent and
child make the same visits inside $I$ until the child
returns $R$.
The parent therefore reaches $(\sigma,p,s)$ immediately
before its new-visit check.
At that point, exactly $r-1$ visits to $p$ have occurred.
No earlier new $r$-th visit in $I$ can have begun at or
after $D$, since the child would have stopped at that
earlier visit.
This proves statement \textup{i}.

\emph{Proof of statement \textup{ii}.}
Suppose $R$ is not an $\mathsf{OVERFLOW}$ report at
a position in $I$.
If $v\ne\bot$, then
\stepref{step:finish-new-visit-report} rejects $R$.
Since $\textbf{FinishAndCheck}$ accepts, we have $v=\bot$.
It follows that $\theta$ is a return report,
$\mathcal H_Y=\varnothing$, and no bit requirement
is added.

The requirements from $\mathcal H_D$ have already
been checked and end at time $t_D$.
Thus, at time $t_D$ and thereafter, the parent and
child perform the same crossing checks and the same
bit checks at positions in $I$.
For a crossing failure, both select the same cut
and record for the failure identifier, since they
consider the cuts in the same order.
For a bit failure, both report the same failed bit
requirement.
The normal-termination rule examines the same unused
crossing records and returns the same report.
The boundary rules also use the same unused records
at boundaries inherited from $\mathcal C$.

The transition and visit cases established above
therefore show that both executions continue in the
same way until they return the same report or the
child stops before a new $r$-th visit in $I$.
If the head enters $J\setminus I$, both return the
same overflow report, as shown above.

The child cannot have stopped before a new $r$-th visit
in $I$, because its report $R$ is not an overflow
report at a position in $I$.
Therefore, $R$ is exactly the parent's next return
report at or after $D$.
This report is produced before any new $r$-th visit
in $I$ begins.
This proves statement \textup{ii}.

\emph{Proof of statement \textup{iii}.}
If $v=\bot$, no visit is completed, so there is
nothing to prove.
Suppose $v\ne\bot$, and let $s$ be the time recorded
in the child's report.
By \stepref{step:finish-new-visit-report}, we have
$s\ge t_D$.

The records supplied by $\mathcal H_D$ have timestamps
less than $t_D$.
The counters restored for completion are therefore
past all these records.
The corresponding requirements end at $t_D$, so
they impose no further crossing checks.

The completion uses the bit $\beta$ at $v$.
Thus the added bit requirement $(v,\beta)$ succeeds.
If $\theta$ is a return report, then
$\mathcal H_Y=\varnothing$, so there are no other
added requirements to check.
If $\theta=\mathsf{CONTINUE}$, a failure caused by
$\mathcal H_Y$ would produce a $\mathsf{FAIL}$ report.
But \stepref{step:finish-compare} requires a
$\mathsf{CONTINUE}$ tuple and rejects such a report.
Since $\textbf{FinishAndCheck}$ accepts, this failure
does not occur.
This proves statement \textup{iii}.

       \item \label{step:verification-completion}
\emph{Every field of $Y$ is the canonical one.}

If $v=\bot$,
\stepref{step:finish-no-new-visit} of
$\textbf{FinishAndCheck}$ excludes an overflow report
at a position in $I$.
By statement \textup{(ii)} of the preceding item,
$R$ is the parent's next return report at or after $D$,
produced before a new $r$-th visit in $I$ begins.

If $v\ne\bot$,
\stepref{step:finish-new-visit-report} of
$\textbf{FinishAndCheck}$ requires an overflow report
at position $v$.
By statement \textup{(i)} of the preceding item,
this report occurs immediately before the parent's
first new $r$-th visit in $I$ at or after $D$.
Since $r\ge2$, the child has already made $r-1$
visits to $v$ and checked $(v,\beta)$ without
returning a failure, so $\beta=(z_I)_v$.
By \Cref{lem:boundary-history-invariant}, the
counters restored in
\stepref{step:finish-new-visit-counters} of
$\textbf{FinishAndCheck}$ equal the child's counters
when it returns this overflow report.
By statement \textup{(iii)} of the preceding item,
the checks added from $D$ and $Y$ cause no failure
during completion.
The completion therefore performs the parent's
allowed $r$-th visit with the correct bit and
remaining requirements.

The checks of the crossing requirements and the
normal-termination rule are applied before a
departure is classified.
If they produce no report,
\stepref{step:finish-new-visit-result} of
$\textbf{FinishAndCheck}$ sets $R'$ to a
$\mathsf{CONTINUE}$ tuple for a departure within
$I$, before checking the new visit at the destination.
This is exactly where
item~\ref{step:iteration} of
\Cref{def:capped-execution} records the next checkpoint.
For entry into $J\setminus I$, the adjacent
destination has already completed $r$ visits, so
both procedures return overflow.
Upon leaving $J$, both use the inherited boundary
history and apply the crossing and normal-termination
checks prescribed in item~\ref{step:iteration} of
\Cref{def:capped-execution}.
If these rules return no report, the supplied entrance
returns to $v$; both then return overflow before
a $(r+1)$-st visit.
Thus, in an accepting test, $R'$ describes the
parent's next checkpoint or return report.

Finally, \stepref{step:finish-compare} of
$\textbf{FinishAndCheck}$ verifies this outcome and
its time.
For a continuation, it also verifies the departure
direction and every history in $\mathcal H_Y$;
for a return report, \Cref{def:candidate} fixes
$d=\bot$ and $\mathcal H_Y=\varnothing$.
Hence all fields agree with the construction in
item~\ref{step:canonical-candidate-fields} of the
proof of \Cref{lem:canonical-candidate}.
Therefore $Y=g_D(z_I)$, proving the forward implication.

    \end{enumerate}

    \item \emph{$Y=g_D(z_I)$ implies acceptance.}
    Suppose $Y=g_D(z_I)$. Since $Y$ is a candidate rather
    than $\bot$, the definition of $g_D$ gives $z_I\in E_D$.
    Choose an extension $w\in A_{j,D}$ with $w|_I=z_I$. By
    \Cref{lem:canonical-candidate}, the next canonical
    candidate of $\textbf{Exec}_r(\mathcal C,w)$ is $Y$.
    The first part of this proof gives $\mathcal C'\ne\bot$.
    We show directly that each part of
    $\textbf{FinishAndCheck}$ succeeds.
    \begin{enumerate}
       \item \emph{The reach-checkpoint check succeeds.}
If $D=D_0$, then
\stepref{step:finish-reach-checkpoint} of
$\textbf{FinishAndCheck}$ is skipped. Assume
$D\ne D_0$ and consider
$\textbf{Exec}_{r-1}(\mathcal C',z_I)$ and
$\textbf{Exec}_r(\mathcal C,w)$. Since $w|_I=z_I$,
these executions read the same bit at every position
of $I$. By the choice of the initial data in
$\textbf{CreateConstraint}$, the execution under
$\mathcal C'$ starts at the initial triple of
$\mathcal C$ if its initial position lies in $I$,
and otherwise immediately after the first entrance
into $I$ recorded by $D$. Its boundary histories
supply the configurations and times at which the
execution under $\mathcal C$ reenters $I$ up to
time $t_D$. Thus, between these prescribed entrances
and exits, the two executions apply the same
transitions of $M$ to the same bits.

No position of $I$ begins its $r$-th visit before
$D$. Otherwise, that visit would end before the
entrance recorded at $D$, and the checkpoint update
would remove that position from the retained interval,
or the execution would stop before reaching $D$.
Hence the visit bound $r-1$ does not stop
$\textbf{Exec}_{r-1}(\mathcal C',z_I)$ before $D$.
The inherited checks pass before $D$ because
$\textbf{Exec}_r(\mathcal C,w)$ reaches $D$.
The histories added from $D$ and $Y$ are taken from
this execution, and the added bit, when present,
equals $(z_I)_v$, so the added requirements also
cause no failure before $D$.
If $\textbf{Exec}_{r-1}(\mathcal C',z_I)$ stops
exactly at time $t_D$, it is at the configuration
and position recorded by $D$ and returns either
internal overflow before the $r$-th visit or a
failure of an inherited bit check at that position.
Both reports are permitted by
\stepref{step:finish-reach-checkpoint} of
$\textbf{FinishAndCheck}$. Thus that step succeeds.

        \item \emph{The completion produces the actual next outcome.}
        The preceding item establishes agreement through
        $D$. Every requirement added from $D$ or $Y$
        prescribes an actual history or bit of the parent
        execution up to the proposed outcome.
        Thus these requirements cause no failure during
        this portion of the execution.
       
        We repeat the comparison in
item~\ref{step:verification-transition-comparison}
for the parent execution
$\textbf{Exec}_r(\mathcal C,w)$ and the child execution
$\textbf{Exec}_{r-1}(\mathcal C',z_I)$, where
$w\in A_{j,D}$ is the extension chosen above and
$w|_I=z_I$.
The executions agree through $D$, read the same bits in
$I$, and impose the same remaining requirements inherited
from $\mathcal C$. The additional requirements agree with
the parent execution because $Y=g_D(z_I)$.
Consequently, each transition inside $I$ and each supplied
entrance produces the same configuration, position, and
time in both executions, until one of the following
events occurs.
If the parent reaches its first new $r$-th visit in $I$,
the child returns internal overflow immediately before
that visit, at the same configuration, position, and time.
If the parent returns a report before such a visit,
the child returns the same report, including its
failure identifier when applicable.
This comparison uses the correctness of the requirements
specified by $Y$, without assuming that
$\textbf{FinishAndCheck}$ accepts.
        If $v=\bot$, the parent produces its next return
        report before beginning a new $r$-th visit in $I$.
        The case without an internal child overflow gives
        exactly the same return report, so
        \stepref{step:finish-no-new-visit} of
        $\textbf{FinishAndCheck}$ sets $R'=R$ without
        rejecting.
        If $v\ne\bot$, the child returns internal overflow
        immediately before the parent's first new $r$-th
        visit to $v$, at a time between $t_D$ and $t_Y$.
        Thus \stepref{step:finish-new-visit-report} of
        $\textbf{FinishAndCheck}$ succeeds.
        The restored counters are correct by
        \Cref{lem:boundary-history-invariant}, and
        $\beta=(z_I)_v$ by the construction of the canonical
        candidate. Hence each transition of the completion
        uses the parent's bit at $v$ and its remaining
        requirements. The departure cases in
        item~\ref{step:verification-completion} of this
        proof then give the parent's next checkpoint or
        return report. No added requirement fails, so none
        of these operations rejects.

        \item \emph{The final comparison succeeds.}
        The outcome just obtained occurs at $t_Y$ and
        agrees with $\theta$, since both are taken from
        the same parent execution.
        For a continuation, $d$ is its actual departure
        direction and $\mathcal H_Y$ consists of its
        actual crossing histories through the new
        checkpoint; these records have been checked during
        the child execution and completion, including any
        initial entrance checked by
        $\textbf{CreateConstraint}$.
        For a return report, all fields of $R'$ equal
        those of $\theta$.
        Hence \stepref{step:finish-compare} of
        $\textbf{FinishAndCheck}$ succeeds, and
        \[
           \textbf{FinishAndCheck}
           (\mathcal C,D,Y,\mathcal C',R)=1.
        \]
        This proves the reverse implication.
    \end{enumerate}
\end{enumerate}
Together, the last two parts prove
\eqref{eq:complete-test-characterization}.
\end{proof}

We now combine the conditional distribution above with the
exact candidate test to prove correctness by induction on the
visit cap.

\begin{lemma}[Exact Sampling]
\label{lem:exact-sampling}
For every $1\le r\le k$ and every
$\mathcal C\in\mathfrak C_r$ on $J$,
$\textbf{Sampler}_r(\mathcal C)$ terminates almost surely
and satisfies
\begin{equation}
 \mathcal L\bigl(\textbf{Sampler}_r(\mathcal C)\bigr)
 =\mathcal L\bigl(\textbf{Exec}_r(\mathcal C,U_J)\bigr).
 \label{eq:exact-sampling-law}
\end{equation}
\end{lemma}

\begin{proof}
\noindent\emph{Induction on the visit cap.}
We prove the statement for all $\mathcal C\in\mathfrak C_r$
simultaneously by induction on $r$.
For $r=1$, $\textbf{Sampler}_1(\mathcal C)$ invokes
$\textbf{DirectSample}(\mathcal C)$.
When $\textbf{DirectSample}$ first reads a position $v$,
it draws a fresh uniform bit $B_v$ and uses $B_v$ for
every read at $v$ during that visit.
In $\textbf{Exec}_1(\mathcal C,U_J)$, the random-tape
assignment is chosen uniformly at the start.
Conditioned on the preceding execution, the bit at a
previously unread position $v$ is still uniform on
$\{0,1\}$. Thus the two procedures use the same
conditional distribution for the bit first read at $v$,
and both reuse that bit throughout the visit.
An attempted second visit
returns overflow before reading the bit again, unless a
preceding crossing check or normal-termination rule has
already returned a report. 
Consequently, $\textbf{DirectSample}$ may discard $B_v$
when the random head leaves $v$: any attempted second
visit to $v$ returns a report before the bit at $v$
is read again.
The two procedures apply the same checks, boundary rules,
and stopping rules, so their return-report distributions
agree. Simulated time strictly increases at each transition
or supplied entrance and is bounded by $T$, which guarantees
termination.

Fix $r\ge2$.
Assume that, for every constraint
$\mathcal C'\in\mathfrak C_{r-1}$ with interval $I$,
$\textbf{Sampler}_{r-1}(\mathcal C')$ terminates
almost surely and satisfies
\[
 \mathcal L\bigl(\textbf{Sampler}_{r-1}(\mathcal C')\bigr)
 =
 \mathcal L\bigl(\textbf{Exec}_{r-1}(\mathcal C',U_I)\bigr).
\]
We prove the corresponding assertions for every
$\mathcal C\in\mathfrak C_r$.
Fix such a constraint $\mathcal C$.

\medskip
\noindent\emph{The conditional distribution of the next outcome.}
Fix a stage $j$ and a checkpoint $D$ reachable at that stage,
and write $I=I_D$. Use the map $g_D$ defined by
\Cref{lem:canonical-candidate}. For each candidate $y$, define
\[
 a_{D,y}:=\Pr[g_D(U_I)=y],
\]
and for each possible next checkpoint or return report $o$,
define
\[
 K_j(D,o):=\Pr[\pi(g_D(U_I))=o\mid U_I\in E_D].
\]
By \Cref{lem:boundary-factorization}, conditioned on
$D_j=D$, the assignment $U_I$ is uniform on $E_D$.
By \Cref{lem:canonical-candidate}, its next outcome is
$\pi(g_D(U_I))$, regardless of which compatible exterior
assignment produced $D$. Hence
\begin{equation}
 K_j(D,o)=\Pr[O_j=o\mid D_j=D].
 \label{eq:checkpoint-transition-kernel}
\end{equation}

The same statement remains true after conditioning on the
earlier checkpoints. To prove this, fix a positive-probability prefix $p=(d_0,\ldots,d_j)$
of the checkpoint sequence, with $d_j=D$. Let $A_p$ be the
set of assignments producing this prefix, and let $F_p$ be
its projection onto $J\setminus I$. 

We claim that
\[
 A_p=\{z_J:z_J|_I\in E_D,\ z_J|_{J\setminus I}\in F_p\}.
\]
The forward inclusion follows from $A_p\subseteq A_{j,D}$.
For the reverse inclusion, choose $u\in E_D$ and $u'\in F_p$.
There are witnesses $z\in A_{j,D}$ and $\widetilde z\in A_p$
with $z|_I=u$ and $\widetilde z|_{J\setminus I}=u'$.
Define $w\in\{0,1\}^J$ by
\[
 w|_I=u,
 \qquad
 w|_{J\setminus I}=u'.
\]
Since $\widetilde z\in A_p\subseteq A_{j,D}$,
\cref{item:2} of the proof of \Cref{lem:boundary-factorization}
applies to $z$ and $\widetilde z$.
It establishes that $w\in A_{j,D}$ and that
$\textbf{Exec}_r(\mathcal C,w)$ and
$\textbf{Exec}_r(\mathcal C,\widetilde z)$ have identical
execution portions in $J\setminus I$ through time $t_D$.
That item also establishes that no position of $I$
completes its $r$-th visit before $D$.
Every earlier noninitial
checkpoint is therefore formed by a departure ending an
$r$-th visit outside $I$, at the same time and with the
same retained interval as in the execution on $\widetilde z$. Moreover,
each earlier retained interval contains $I$. Its boundary
cuts therefore either lie entirely outside $I$, where the
execution is unchanged, or coincide with a boundary of $I$,
where the crossings are fixed by $\mathcal H_D$. All earlier
checkpoint histories are consequently unchanged. 
Thus the checkpoint sequence produced by
$\textbf{Exec}_r(\mathcal C,w)$ begins with
$d_0,\ldots,d_j$.
By the definition of $A_p$, we have $w\in A_p$.
This proves the reverse inclusion.

Each assignment in $E_D$ consequently has exactly $|F_p|$
extensions in $A_p$. As in the final step of
\Cref{lem:boundary-factorization}, this gives
$\mathcal L(U_I\mid U_J\in A_p)=\operatorname{Unif}(E_D)$.
The canonical-candidate lemma applies to every such extension,
so the next outcome is still $\pi(g_D(U_I))$. Therefore
\begin{equation}
 \Pr[O_j=o\mid D_0=d_0,\ldots,D_j=d_j]
 =K_j(D,o).
 \label{eq:checkpoint-markov-property}
\end{equation}
This is the Markov property of the checkpoint process.

We will use \eqref{eq:checkpoint-markov-property} to compare
the probabilities of complete checkpoint sequences.

\medskip
\noindent\emph{The distribution of an accepted trial.}
Since $g_D$ is a candidate exactly on $E_D$, the definitions
of $a_{D,y}$ and $K_j$ give
\[
 q_D K_j(D,o)=\sum_{y:\,\pi(y)=o}a_{D,y},
 \qquad
 \sum_y a_{D,y}=q_D>0.
\]

One trial of $\textbf{Sampler}_r$ at $D$ has the form
\[
\begin{aligned}
Y&\gets\textbf{NextGuess}_r(\mathcal C,D),\\
\mathcal C'&\gets
   \textbf{CreateConstraint}(\mathcal C,D,Y),\\
R&\gets\textbf{Sampler}_{r-1}(\mathcal C'),\\
\mathsf{ok}&\gets
   \textbf{FinishAndCheck}(\mathcal C,D,Y,\mathcal C',R).
\end{aligned}
\]
Here $Y$ is the random candidate produced by
$\textbf{NextGuess}$, $\mathcal C'$ is the resulting child
constraint, and $R$ is the child report.
If either of the first two lines returns $\bot$, the remaining
lines are skipped and we set $\mathsf{ok}=0$. Otherwise, the
trial is accepted exactly when $\mathsf{ok}=1$.

For a fixed candidate $y$ such that
$\mathcal C_y:=\textbf{CreateConstraint}(\mathcal C,D,y)$ is
not $\bot$, the definition of the families gives
$\mathcal C_y\in\mathfrak C_{r-1}$. The induction hypothesis
applied to the child call in
\stepref{step:sample-child} of $\textbf{Sampler}_r$ gives
\[
 \mathcal L\bigl(\textbf{Sampler}_{r-1}(\mathcal C_y)\bigr)
 =\mathcal L\bigl(\textbf{Exec}_{r-1}(\mathcal C_y,U_I)\bigr).
\]
Applying \Cref{lem:outcome-test} gives
\[
 \Pr\!\left[
  \textbf{FinishAndCheck}\bigl(\mathcal C,D,y,\mathcal C_y,
  \textbf{Exec}_{r-1}(\mathcal C_y,U_I)\bigr)=1
 \right]
 =\Pr[g_D(U_I)=y]
 =a_{D,y}.
\]
If $a_{D,y}>0$, some $z_I\in E_D$ satisfies $g_D(z_I)=y$.
The first assertion of \Cref{lem:outcome-test} then gives
$\mathcal C_y\ne\bot$. Thus a candidate for which the
construction returns $\bot$ has $a_{D,y}=0$. Malformed
encodings are also rejected.

Each candidate has a unique canonical encoding in the
encoding space used by $\textbf{NextGuess}_r$.
Since that procedure samples uniformly from its $Q_r$
encodings, one trial satisfies
\[
 \Pr[\mathsf{ok}=1\text{ and }\pi(Y)=o]=\frac{1}{Q_r}\sum_{y:\,\pi(y)=o}a_{D,y}=\frac{q_D}{Q_r}K_j(D,o),
\]
\[
 \Pr[\mathsf{ok}=1]=\frac{q_D}{Q_r}.
\]
Write $p_D=q_D/Q_r>0$. For each $h\ge1$, the probability
that the first accepted trial is trial $h$ and its outcome
is $o$ equals
\[
 (1-p_D)^{h-1}p_D K_j(D,o).
\]
Summing over $h$ shows that the accepted outcome has
distribution $K_j(D,\cdot)$. The probability of rejecting
the first $h$ trials is $(1-p_D)^h$, which tends to zero.
Each child call terminates almost surely by induction,
so each reached stage terminates almost surely.

\medskip
\noindent\emph{Termination and equality of the output distributions.}
For a sequence of checkpoints $p=(d_0,\ldots,d_j)$,
let $\widehat A_p$ be the event that
$\textbf{Sampler}_r(\mathcal C)$ records
$d_0,\ldots,d_j$ as its first $j+1$ checkpoints,
in this order.
Recall that $A_p$ is the set of assignments for which
$\textbf{Exec}_r(\mathcal C,z_J)$ records these same
checkpoints in this order.

We prove by induction on $j$ that, for every such $p$,
\[
 \Pr[\widehat A_p]=\Pr[U_J\in A_p].
\]
For $j=0$, both procedures start at $D_0$ with
probability $1$, so the equality holds.

Suppose the equality holds for every sequence of
$j+1$ checkpoints.
Fix $p=(d_0,\ldots,d_j)$ such that
$\Pr[\widehat A_p]>0$, and put $D=d_j$ and $I=I_D$.
The induction hypothesis on $j$ gives
\[
 \Pr[U_J\in A_p]=\Pr[\widehat A_p]>0.
\]
Consequently, $A_p$ is nonempty.
Any assignment in $A_p$ makes the capped execution
reach $D$ at stage $j$.
Thus $D$ is reachable at that stage, and every
non-$\bot$ child constraint constructed at $D$
belongs to $\mathfrak C_{r-1}$.

Every child call at $D$ has the form
$\textbf{Sampler}_{r-1}(\mathcal C')$ with
$\mathcal C'\in\mathfrak C_{r-1}$.
Recall the assumption stated at the beginning of this proof,
each such call terminates almost surely and returns
a report with the same distribution as
$\textbf{Exec}_{r-1}(\mathcal C',U_I)$.
The preceding trial calculation shows that the trials
at $D$ eventually accept almost surely and that the
accepted outcome has distribution $K_j(D,\cdot)$.
This calculation remains valid conditioned on
$\widehat A_p$: the current checkpoint is $D$, and
all subsequent trials use fresh independent coins.
For the capped execution,
\eqref{eq:checkpoint-markov-property} gives the same
conditional distribution for its next outcome,
conditioned on $U_J\in A_p$.

For any next checkpoint $d_{j+1}$, write
$p'=(d_0,\ldots,d_j,d_{j+1})$.
It follows that
\[
\begin{aligned}
 \Pr[\widehat A_{p'}]
 &=\Pr[\widehat A_p]K_j(D,d_{j+1})\\
 &=\Pr[U_J\in A_p]K_j(D,d_{j+1})\\
 &=\Pr[U_J\in A_{p'}].
\end{aligned}
\]
If $\Pr[\widehat A_p]=0$, the induction hypothesis
gives $\Pr[U_J\in A_p]=0$, and every extension of $p$
has probability zero in both procedures.
This completes the induction on $j$.

The same calculation applies when the next outcome
is a return report $\rho$.
For every positive-probability prefix
$p=(d_0,\ldots,d_j)$, the probability that the sampler
records these checkpoints and then returns $\rho$
before recording another checkpoint is
\[
 \Pr[\widehat A_p]K_j(d_j,\rho).
\]
For the capped execution, the corresponding probability is
\[
 \Pr[U_J\in A_p]K_j(d_j,\rho).
\]
These probabilities are equal.

Each accepted continuation strictly decreases the size
of the current nonempty interval, so the sampler reaches
at most $|J|$ stages.
At every prefix having positive probability, the trials
at the last checkpoint terminate almost surely, as
proved above.
There are finitely many possible prefixes, so the
probability that any reached stage fails to terminate
is zero.
Hence $\textbf{Sampler}_r(\mathcal C)$ terminates
almost surely.

For a fixed report $\rho$, the events specifying all
checkpoints recorded before returning $\rho$ are
pairwise disjoint and together constitute the event
that the output is $\rho$.
Summing their equal probabilities therefore gives
\[
 \Pr[\textbf{Sampler}_r(\mathcal C)=\rho]
 =
 \Pr[\textbf{Exec}_r(\mathcal C,U_J)=\rho].
\]
This proves \eqref{eq:exact-sampling-law} and completes
the induction on $r$.
\end{proof}

\subsection{Space, Expected Time and Truncation}
\label{subsec:conclusion}

We first analyze the ideal sampler before applying the global
time cutoff.

\begin{lemma}[Space and Expected Time]
\label{lem:ideal-sampler-complexity}
Let $\mathcal C_0$ be the initial constraint from
\Cref{lem:initial-execution}. Then
$\textbf{Sampler}_k(\mathcal C_0)$ uses $O(k^2\mathfrak{b})$ space and has
expected running time $2^{O(k^2\mathfrak{b})}$.
\end{lemma}

Recall that
\[
 \mathfrak{b}:=C_0\bigl(S+\log(n+T+2)\bigr),
\]
where $C_0$ is a sufficiently large constant depending only
on $M$. Thus $\mathfrak{b}$ bits suffice to encode a
configuration, a random-head position, and a timestamp,
as fixed at the beginning of \Cref{sec:fresh-coin}.

\begin{proof}

At visit cap $r$, the current checkpoint $D$ and candidate
$Y$ each contain $O(r)$ crossing records of $O(\mathfrak{b})$ bits.
Each constraint construction adds only constantly many
crossing requirements and at most one bit requirement.
Thus a frame has $O(k)$ inherited requirements. Store their
lists by references to the enclosing frames: merging selects
one existing list, and deleting a consumed prefix changes
only its starting index. Each reference, cutoff, and history
counter uses $O(\mathfrak{b})$ bits. These references remain valid while
the enclosing call is suspended. The frame therefore uses
$O(r\mathfrak{b})$ bits for $D,Y$ and $O(k\mathfrak{b})$ bits for inherited
requirements, counters, configurations, and reports.
Rejected trials reuse this storage.
The work-space guard specified at the beginning of
\Cref{sec:fresh-coin} ensures that the same configuration
bound applies even when a trial uses an incorrect guess.

In $\textbf{DirectSample}$, the positions visited inside its
interval form an interval, because the head moves locally
and every supplied outside excursion returns to the endpoint
from which it exited. Store the two endpoints of the visited
interval and the bit used during the current visit. An entry
into a previously visited position is then detected using
$O(\log(T+2))$ space, without storing a visit counter for
every position. The transition checks and normal-termination
rule retain their prescribed precedence over this test.
Consequently, the total space is
\[
 \sum_{r=1}^k O((r+k)\mathfrak{b})=O(k^2\mathfrak{b}).
\]

For the expected running time, use the families
$\mathfrak C_r$ defined in \Cref{subsec:sample-correctness}
and define
\[
 A_r:=\sup_{\mathcal C\in\mathfrak C_r}
 \mathbb E\!\left[
   \operatorname{Time}\bigl(\textbf{Sampler}_r(\mathcal C)\bigr)
 \right].
\]
The supremum is also taken over the representations of these
constraints by enclosing frames. The expectation includes
the fresh coins used by the call and all of its descendants.
Thus $A_{r-1}$ uniformly bounds the
recursive part of every trial at visit cap $r$.

Let $H$ bound the deterministic work in one trial outside its
recursive call. A direct implementation may rescan at most $k$
active frames and simulate at most $T$ transitions, so we may
take
\[
 H:=(n+T+k+\mathfrak{b}+2)^{O(k)}.
\]
Since $k\le T+1$ and $\mathfrak{b}=\Omega(\log(n+T+2))$, we have
$H=2^{O(k\mathfrak{b})}$. The base procedure $\textbf{DirectSample}$
makes no recursive call, so $A_1\le H$.

A candidate encoding has $O(r\mathfrak{b})$ bits, so the code-space size
fixed in \textbf{NextGuess} satisfies $Q_r=2^{O(r\mathfrak{b})}$. A
checkpoint also has $O(r\mathfrak{b})$ bits. Let $N_r$ denote the number
of possible checkpoints at a fixed stage; then
$N_r=2^{O(r\mathfrak{b})}$ as well.

We derive the recurrence for $A_r$ by induction, with
$A_1\le H$ as the base case. Assume $A_{r-1}<\infty$.
Fix a stage $j$ and a
reachable checkpoint $D$. By the calculation in the proof of
\Cref{lem:exact-sampling}, one trial succeeds with probability
$q_D/Q_r$. Its expected cost is at most $A_{r-1}+H$.
To bound the total cost, index the trials by $h\ge1$.
Trial $h$ is reached with probability
$(1-q_D/Q_r)^{h-1}$. Its fresh coins are independent of
the preceding trials, so its expected cost, conditioned
on being reached, is at most $A_{r-1}+H$.
Summing over $h$ gives expected cost at most
\[
 \frac{Q_r}{q_D}(A_{r-1}+H)
\]
at $D$. The probability that stage $j$ reaches $D$ is
$\Pr[U_J\in A_{j,D}]$. By
\Cref{lem:boundary-factorization}, this probability is at most
$q_D$. Hence the expected cost of stage $j$, averaged over its
possible checkpoints, is at most
\[
 \sum_{D:\Pr[U_J\in A_{j,D}]>0}\Pr[U_J\in A_{j,D}]\frac{Q_r}{q_D}(A_{r-1}+H)
 \le N_rQ_r(A_{r-1}+H).
\]
There are at most $T+1$ stages in one call. Therefore, for an
absolute constant $C$,
\[
 A_r\le C(T+1)N_rQ_r(A_{r-1}+H).
\]
Since $N_rQ_r=2^{O(r\mathfrak{b})}$ and $\log(T+1)=O(\mathfrak{b})$, iterating this
recurrence from $A_1\le H$ gives
\[
 \log(A_k+H)
 \le O(k\mathfrak{b})+\sum_{r=2}^k O(r\mathfrak{b})
 =O(k^2\mathfrak{b}).
\]
Taking $r=k$ and the initial constraint $\mathcal C_0$ gives
the claimed expected-time bound. Together with the space bound
above, this proves the lemma.
\end{proof}

We finally truncate the running time of the ideal sampler to
obtain a worst-case-time simulation and prove
\Cref{thm:local-read-simulation}.

\begin{proof}[Proof of \Cref{thm:local-read-simulation}]
Fix an input $x$ of length $n$, and let $M$ be a locally
read-$k$ machine running in time $T$ and space $S$. Let
$\mathcal C_0$ be the initial constraint from
\Cref{lem:initial-execution}, and let $U_{J_0}$ be uniform
on $\{0,1\}^{J_0}$. By \Cref{lem:initial-execution},
$\textbf{Exec}_k(\mathcal{C}_0, U_{J_0})$ returns a
$\mathsf{HALT}$ report whose configuration contains the
output of $M$ on the same random tape.
Furthermore, by \Cref{lem:exact-sampling},
\[
 \mathcal L\bigl(\textbf{Sampler}_k(\mathcal C_0)\bigr)
 =
 \mathcal L\bigl(\textbf{Exec}_k(\mathcal C_0,U_{J_0})\bigr).
\]
Thus, returning the output recorded in the ideal sampler's
halting configuration gives exactly the output distribution
of $M$, using only fresh random bits.

By \Cref{lem:ideal-sampler-complexity}, choose a sufficiently
large constant $C_3$ so that $A:=2^{C_3k^2\mathfrak{b}}$ bounds the
expected running time of the ideal sampler, and
put $E:=\left\lceil\log_2(1/\varepsilon)\right\rceil$.

Define $M'$ by running the ideal sampler for at most $A2^E$
steps. If it returns in time, output the value recorded in
its halting configuration; otherwise, output a fixed value.
Markov's inequality gives
\[
 \Pr[\text{$M'$ times out}]
 \le \frac{A}{A2^E}
 \le \varepsilon.
\]
Couple $M'$ and the ideal sampler using the same fresh random
bits. Their outputs agree unless $M'$ times out, so their
output distributions have total variation distance at most
$\varepsilon$. Since the ideal sampler has the same output
distribution as $M$, the same bound holds between the
output distributions of $M'$ and $M$.

By \Cref{lem:ideal-sampler-complexity}, the ideal sampler uses
$O(k^2\mathfrak{b})$ space. A counter for the
time limit uses
\[
 O\bigl(\log(A2^E)\bigr)=O(k^2\mathfrak{b}+E)
\]
additional bits and can be maintained in amortized constant
time per simulated step. Since $2^E\le 2/\varepsilon$, the
worst-case running time of $M'$ is
\[
 O(A2^E)
 =
 \frac{1}{\varepsilon}\cdot 2^{O(k^2\mathfrak{b})}.
\]
Substituting
$\mathfrak{b}=O(S+\log(n+T+2))$ and
$E=O(\log(1/\varepsilon))$ gives space
\[
 O\bigl(k^2(S+\log(n+T+2))
       +\log(1/\varepsilon)\bigr)
\]
and worst-case time
\[
 \frac{1}{\varepsilon}\cdot
 2^{O(k^2(S+\log(n+T+2)))}.
\]
These bounds prove the theorem.
\end{proof}

To deduce \Cref{cor:local-read-complexity}, we must verify
that the time bound $T(n)=2^{O(S(n))}$ can be initialized
within the claimed resources, since logspace constructibility
of $S$ does not necessarily imply logspace constructibility
of $T$.
\begin{proof}[Proof of \Cref{cor:local-read-complexity}]
For the read-once class $\BPSPACE(S)$, we use the same
$2^{O(S)}$ time convention as in the definition of
$\LocalRead_k$-$\BPSPACE(S)$.
Fix a constant $k\ge1$ and a logspace constructible function
$S(n)\ge\log n$.
The inclusion
$\BPSPACE(S)\subseteq\LocalRead_k$-$\BPSPACE(S)$
follows because a read-once random tape is locally read-$1$.

For the reverse inclusion, let $M$ witness membership in
$\LocalRead_k$-$\BPSPACE(S)$.
Choose a constant integer $c\ge1$ such that
$T(n):=2^{cS(n)}$ bounds its running time.
Apply the construction in the proof of
\Cref{thm:local-read-simulation} with this time bound and
$\varepsilon=1/12$.
For this choice of $T$, the required parameter initialization
can be performed in $O(S(n))$ space and $2^{O(S(n))}$ time:
compute $S(n)$ using its logspace construction, form the
binary representation of $T(n)$, which has $cS(n)+1$ bits,
and initialize the simulation's clocks.
The same bounds accommodate the parameter computations,
since $S(n)\ge\log n$.
Thus the construction is uniform within the claimed
resources, without requiring a logspace construction of
$T(n)$.

Since $k$ is constant and
$\log(n+T(n)+2)=O(S(n))$, the resulting read-once machine
uses $O(S(n))$ space and $2^{O(S(n))}$ time.
Its acceptance probability is at least $7/12$ on yes-instances
and at most $5/12$ on no-instances.
A constant number of independent repetitions followed by
a majority vote reduces the error to at most $1/3$,
preserving both resource bounds.
This proves the reverse inclusion and hence the equality.
Taking $S(n)=\Theta(\log n)$ gives
$\LocalRead_k$-$\BPL=\BPL$.
\end{proof}

\section*{Statement of AI use}

 We formulated the research topics and questions of this work. We used ChatGPT (GPT-6 Astra) to assist with proof development and proof checking, making substantial use of the tool for both tasks. The tool materially affected the proofs throughout all technical sections of the paper. The manuscript was written by the authors with assistance from the same tool. The authors take full responsibility for the correctness of the work.
 
\bibliographystyle{alpha}
\bibliography{refer}

\appendix

\end{document}